\documentclass[printer]{tADP2e}
\usepackage{rotating}

\begin{document}
\doi{10.1080/00018730xxxxxxxxxxxx}
 \issn{1460-6976}
\issnp{0001-8732} \jvol{00} \jnum{00} \jyear{2005} \jmonth{January-February}

\markboth{Karen A. Hallberg}{New Trends in Density Matrix Renormalization}

\title{{\itshape New Trends in Density Matrix Renormalization}}
\author{KAREN A. HALLBERG \\Instituto Balseiro and Centro At\'omico Bariloche,
Comisi\'on Nacional de Energ\'{\i}a At\'omica, 8400 Bariloche, Argentina}  \received{April 2006}

\maketitle

\begin{abstract}
The Density Matrix Renormalization Group (DMRG) has become a powerful
numerical method that can be applied to
low-dimensional strongly correlated fermionic and bosonic systems. It
allows for a very precise calculation of static, dynamic and  thermodynamic
properties.
Its field of applicability has now extended beyond Condensed
Matter, and is successfully used in Quantum Chemistry, Statistical Mechanics, 
Quantum Information Theory, Nuclear and High Energy Physics as well.
In this article, we briefly review the main aspects of the
method and present some of the most relevant applications so as to
give an overview on the scope and possibilities of DMRG. We focus on the
most important extensions of the method such as the calculation of
dynamical properties, the application to
classical systems, finite temperature simulations, phonons and disorder, field
theory, time-dependent properties and the {\it ab initio} calculation of
electronic states in molecules. 
The recent quantum information interpretation, the development of highly accurate
time-dependent algorithms and the possibility of using 
the DMRG as the impurity-solver of the Dynamical Mean Field Method (DMFT) give 
new insights into its present and potential uses.  
We review the numerous very recent applications of these techniques where the DMRG has
shown to be one of the most reliable and versatile methods in modern computational 
physics.\bigskip

 \bigskip
\pagebreak

\centerline{\bf Contents}\bigskip

\noindent\hbox to
\textwidth{\hsize\textwidth\vbox{\noindent\hsize20.1pc
{1.}    Introduction\\
\hspace*{10pt}{1.1.} Motivation\\
{2.}    The DMRG Method\\
\hspace*{10pt}{2.1.} The density matrix projection\\
\hspace*{20pt}{2.1.1} Quantum information interpretation \\
\hspace*{10pt}{2.2.} General remarks\\
\hspace*{10pt}{2.3.} Symmetries\\
{3.}    Applications\\
{4.}    Bosonic degrees of freedom\\
{5.}    Generalized DMRG\\
\hspace*{10pt}{5.1.} Momentum representation\\
\hspace*{10pt}{5.2.} Grains, nuclei and high energy physics\\
\hspace*{10pt}{5.3}  Molecules and Quantum Chemistry\\
{6.}    Higher dimensions ($D>1$)\\
{7.}    Time-dependent analysis\\
{8.}    Zero temperature dynamics\\
\hspace*{10pt}{8.1.}   Lanczos technique\\
\hspace*{10pt}{8.2.}   Correction-vector method\\
{9.}    Classical systems\\
{10.}    Finite temperature DMRG\\
\hspace*{10pt}{10.1.}   Finite temperature dynamics\\
{11.}    Dynamical Mean Field Theory using DMRG\\   
{12.}   Summary and outlook\\
   References\\
   Acknowledgments\\
}}

\end{abstract}

\section{Introduction}

The Density Matrix Renormalization Group (DMRG) was developed by S. White in
1992 \cite{white1} and since then it has proved to be a very powerful method
for low dimensional interacting systems. Its remarkable accuracy can be seen
for example in the spin-1 Heisenberg chain: for a system of hundreds of sites a
precision of $10^{-10}$ for the ground state energy can be achieved. Since then
it has been applied to a great variety of systems and problems including, among
others, spin chains and ladders, fermionic and bosonic systems, disordered
models, impurities, molecules, nanoscopic systems and 2D electrons in high magnetic fields. It
has also been improved substantially in several directions like two dimensional
(2D) classical systems, stochastic models, inclusion of phonons, quantum
chemistry, field theory, finite temperature and the calculation of dynamical
and time-dependent properties. Some calculations have also been performed in 2D
quantum systems. Most of these topics are treated in detail and in a
pedagogical way in the book \cite{book}, where the reader can find an extensive
overview on the foundations of  DMRG and also in a comprehensive review \cite{scholl}.
Recent new developments of the DMRG like the
implementation of very accurate methods for time-dependent problems, the quantum information
and matrix-product perspective and the possibility to combine it with the Dynamical
Mean Field Theory (DMFT), triggered a great activity using these techniques, as seen by the
numerous recent papers published in the last year.  
In this article we will mainly focus on these new developments and applications and hope to
give new insights into its potential uses. In order to achieve this and on behalf of 
coherence
we will also briefly describe the method and its former extensions.

\subsection{Motivation}

When considering finite systems, the exponential growth of degrees of
freedom to be considered imposes an important limitation in numerical
calculations. Several methods have been introduced in order to reduce the
size of the Hilbert space to be able to reach larger systems, such as Monte
Carlo, renormalization group (RG) and DMRG. Each method considers a
particular criterion for keeping the relevant information.

The DMRG was originally developed to overcome the problems that arise in
interacting systems in 1D when standard RG procedures were applied. For example, consider
a block B (a block is a collection of sites) where the Hamiltonian $H_B$ and
end-operators are defined. These traditional methods consist in putting
together two or more blocks (e.g. B-B', which we will call the superblock),
connected using end-operators, in a basis that is a direct product of the
basis of each block, forming $H_{BB^{\prime}}$. This Hamiltonian is then
diagonalized, the superblock is replaced by a new effective block $B_{new}$
formed by a certain number $m$ of lowest-lying eigenstates of $%
H_{BB^{\prime}}$ and the iteration is continued (see Ref.~\cite{white2}).
Although it has been used successfully in certain cases, this procedure, or
similar versions of it, has been applied to several interacting systems with
poor performance. For example, it has been applied to the 1D Hubbard model
keeping $m\simeq 1000$ states and for 16 sites, an error of 5-10\% was obtained
\cite{braychui}. Other results\cite{panchen} were also discouraging. A better
performance was obtained \cite{xiangghering} by adding a single site at a time
rather than doubling the block size. However, there is one case where a similar
version of this method applies very well: the impurity problem represented by the Kondo and Anderson models.
Wilson\cite{wilson} mapped the one-impurity problem onto a one-dimensional
lattice with exponentially decreasing hoppings. The difference with the method
explained above is that in this case, one site (equivalent to an ``onion
shell") is added at each step and, due to the exponential decrease of the
hopping, very accurate results can be obtained. A very recent work on
renormalization group transformations on quantum states based on matrix product
states was performed in Ref. \cite{verstlatorre}.

Returning to the problem of putting several blocks together, the main source of
error comes from the election of eigenstates of $H_{BB^{\prime}}$ as
representative states of a superblock. Since $H_{BB^{\prime}}$ has no
connection to the rest of the lattice, its eigenstates may have unwanted
features (like nodes) at the ends of the block and this can't be improved by
increasing the number of states kept. Based on this consideration, Noack and
White\cite{noackwhite} tried including different boundary conditions and
boundary strengths. This turned out to work well for single particle and
Anderson localization problems but, however, it did not improve the results
significantly for interacting systems. These considerations led to the idea of
taking a larger superblock that includes the blocks $BB^{\prime}$, diagonalize
the Hamiltonian in this large superblock and then somehow project the most
favorable states onto $BB^{\prime}$. Then $BB^{\prime}$ is replaced by
$B_{new}$. In this way, awkward features in the boundary would vanish and a
better representation of the states in the infinite system would be achieved.
White\cite{white1,white2} appealed to Feynman's formulation of the density
matrix as the best description of a part of a quantum mechanical system and,
thus, as the optimal way of projecting the most relevant states onto a
subsystem. In Ref.\cite{gaite}, Gaite shows that, by using an angular
quantization construction of the density matrix, the DMRG is an algorithm that
keeps states with a higher weight near the boundaries in a systematic way. This
is a consequence of the fact that, near the boundaries, there is a
concentration of quantum states, in a similar manner as in the physics of black
holes.

In the following Section we will describe the standard method; in Sect. 3 we
will mention some of the most important applications, describing in Sect. 4 the
treatment of bosonic degrees of freedom. In Sec. 5 we review the most relevant
extensions of the method like the momentum and energy levels representation and
the application to Quantum Chemistry. The improvement to treat systems in
dimensions higher than 1 is mentioned in Sec. 6, while Sect. 7 will be devoted
to the latest and very promising developments to calculate non-equilibrium and time-dependent
properties. In Sect. 8 we deal with 
the calculation of dynamical properties at zero temperature and in Sect. 9
we briefly 
describe how DMRG can be used to calculate physical properties of classical
systems. In connection to this, finite temperature studies were possible within
DMRG and this is explained in Section 10. In the following Sect. 11 we describe how 
the Dynamical Mean Field Theory (DMFT) can profit from the DMRG as an impurity-solver
to obtain spectral properties of correlated systems, opening new posibilities for the 
calculation of more complicated and realistic systems. 
 Finally, we summarize the achievements of the method and look upon
new potential applications and developments.

\section{The DMRG Method}

The DMRG allows for a systematic truncation of the Hilbert space by keeping
the most probable states describing a wave function (\textit{e.g.~}the
ground state) instead of the lowest energy states usually kept in previous
real space renormalization techniques.

The basic idea consists in starting with a small system (\textit{e.g} with $N
$ sites) and then gradually increase its size (to $N+2$, $N+4$,...) until
the desired length is reached. Let us call the collection of $N$ sites the
\textit{universe} and divide it into two parts: the \textit{system} and the
\textit{environment} (see Fig. 1). The Hamiltonian is constructed in the
\textit{universe} and its ground state $|\psi_0\rangle$ is obtained. This is
considered as the state of the \textit{universe} and called the \textit{%
target state}. It has components on the \textit{system} and the \textit{%
environment}. We want to obtain the most relevant states of the \textit{%
system}, i.e., the states of the \textit{system} that have largest weight in
$|\psi_0\rangle$. To obtain this, the \textit{environment} is considered as
a statistical bath and the density matrix\cite{feynman} is used to obtain
the desired information on the \textit{system}. So instead of keeping
eigenstates of the Hamiltonian in the block (\textit{system}), we keep
eigenstates of the density matrix. We will be more explicit below.

A very easy and pedagogical way of understanding the basic functioning of DMRG
is applying it to the calculation of simple quantum problems like one particle
in a tight-binding chain \cite {whitebook,sierraparticle}. In these examples, a
discretized version of the Schr\"odinger equation is used and extremely accurate
results are obtained for the quantum harmonic oscillator, the anharmonic
oscillator and the double-well potential.


\begin{figure}[tbp]
\begin{center}
\epsfxsize=3.0in \epsfysize=1.5in \epsffile{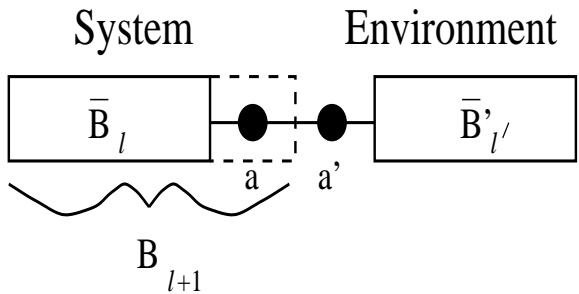}
\end{center}
\caption{A scheme of the superblock (universe) configuration for the DMRG
algorithm\protect\cite{white2}.}
\label{fig1}
\end{figure}

Let's define block [\textbf{B}] as a finite chain with $l$ sites having an
associated Hilbert space with $M$ total amount of states (or $m$ states after a reduction is
performed) where operators are defined (in
particular the Hamiltonian in this finite chain, $H_B$, and the operators at
the ends of the block, useful for linking it to other chains or added
sites). Except for the first iteration, the basis in this block isn't
explicitly known due to previous basis rotations and reductions. The
operators in this basis are matrices and the basis states are characterized
by quantum numbers (like $S^z$, charge or number of particles, etc). We also
define an added block or site as [\textbf{a}] having $n$ states. A general
iteration of the method is described below:

i) Define the Hamiltonian $H_{BB^{\prime}}$ for the superblock (the \textit{%
universe}) formed by putting together two blocks [\textbf{B}] and [\textbf{B'%
}]  and two added sites [\textbf{a}] and [\textbf{a'}] in this way: [\textbf{%
B a a' B' }], where the primes indicate additional blocks. The
primed blocks usually have the same structure as the non-primed ones, but this can vary
(see the finite-system algorithm below). The block [\textbf{B'}] has $M'$ or $m'$
states for the full and reduced spaces, respectively.
In general, blocks [\textbf{B}] and [%
\textbf{B'}] come from the previous iteration. The total Hilbert space of
this superblock is the direct product of the individual spaces corresponding
to each block and the added sites. In practice a quantum number of the
superblock can be fixed (in a spin chain for example one can look at the
total $S^z=0$ subspace), so the total number of states in the superblock is
much smaller than $(mn)^2$. In some cases, as the quantum number of the
superblock consists of the sum of the quantum numbers of the individual
blocks, each block must contain several subspaces (several values of $S^z$
for example). Here periodic boundary conditions can be attached to the ends
and a different block layout should be considered (e.g. [\textbf{B a B' a' }%
]) to avoid connecting blocks [\textbf{B}] and [\textbf{B'}] which takes
longer to converge. The boundary conditions are between [\textbf{a'}] and [%
\textbf{B}]. For closed chains the performance is poorer than for open
boundary conditions \cite{white2,scalapino} (see Fig. 2)

ii) Diagonalize the Hamiltonian $H_{BB^{\prime}}$ to obtain the ground state
$|\psi_0\rangle$ (target state) using Lanczos\cite{lanczos} or Davidson\cite
{davidson} algorithms. Other states could also be kept, such as the first
excited ones: they are all called \textit{target states}.
A faster convergence of Lanczos or Davidson algorithm is achieved by
choosing a good trial vector\cite{cavo,white4}.

iii) Construct the density matrix:
\begin{equation}
\rho_{ii^{\prime}}=\sum_j \psi_{0,ij}\psi_{0,i^{\prime}j}
\end{equation}
on block [\textbf{B a}], where $\psi_{0,ij}=\langle i\otimes j|\psi_0\rangle
$, the states $|i\rangle $ belonging to the Hilbert space of the block [%
\textbf{B a}] and the states $|j\rangle $ to the block [\textbf{B' a'}]. The
density matrix considers the part [\textbf{B a}] as a system and [\textbf{B'
a'}], as a statistical bath. The eigenstates of $\rho$ with the highest
eigenvalues correspond to the most probable states (or equivalently the
states with highest weight) of block [\textbf{B a}] in the ground state of
the whole superblock. These states are kept up to a certain cutoff, keeping
a total of $m$ states per block. The density matrix eigenvalues,
$\omega_{\alpha}$, sum up to
unity and the truncation error, defined as the sum of the density matrix
eigenvalues corresponding to discarded eigenvectors, $\sum_{\alpha=m+1}^{m_{max}}
\omega_{\alpha}=1-sum_{\alpha=1}^{m}\omega_{\alpha}$, 
gives a qualitative indication of the accuracy of the calculation ($m_{max}$
corresponds to the size of the full space of block [\textbf{B a}]).

iv) With these $m$ states a rectangular matrix $O$ is formed and it is used
to change basis and reduce all operators defined in [\textbf{B a}]. This
block [\textbf{B a}] is then renamed as block [\textbf{B$_{new}$}] or simply
[\textbf{B}] (for example, the Hamiltonian in block [\textbf{B a}], $H_{Ba}$%
, is transformed into $H_{B}$ as $H_{B}=O^\dagger H_{Ba} O$).

v) A new block [\textbf{a}] is added (one site in our case) and the new
superblock [\textbf{B a a' B'}] is formed as the direct product of the
states of all the blocks.

vi) This iteration continues until the desired length is achieved. At each
step the length is $N=2l+2$ (if [\textbf{a}] consists of one site).

When more than one target state is used, \textit{i.e} more than one state is
wished to be well described, the density matrix is defined as:
\begin{equation}  \label{eq:pl}
\rho_{ii^{\prime}}=\sum_l p_l \sum_j \phi_{l,ij} \phi_{l,i^{\prime}j}
\end{equation}
where $p_l$ defines the probability of finding the system in the target
state $|\phi_l\rangle $ (not necessarily eigenstates of the Hamiltonian).

\begin{figure}
\begin{center}
\vspace{1cm}
\epsfxsize=3.0in \epsfysize=2.0in \epsffile{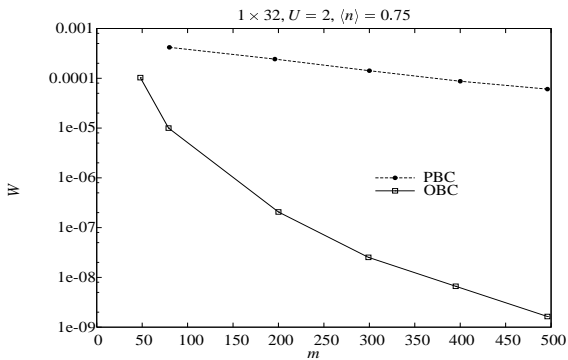}
\end{center}
\caption{Truncation error for the finite-system DMRG as a function of the number of
states kept $m$ (reprinted from \cite{whitebook}, with permission).}
\label{fig2}
\end{figure}

\begin{figure}
\begin{center}
\epsfxsize=3.0in \epsfysize=1.5in \epsffile{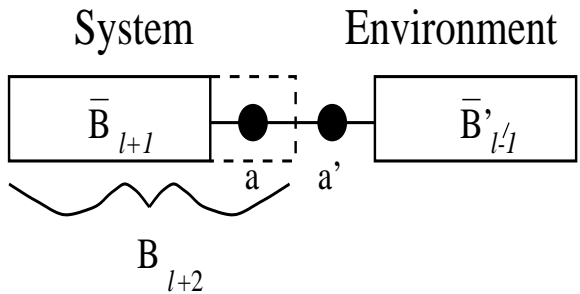}
\end{center}
\caption{One step of the finite-system algorithm, where the total length $N=l+l'+2$
is fixed.}
\label{fig3}
\end{figure}

\begin{figure}[tbp]
\begin{center}
\epsfxsize=3.0in \epsfysize=2.0in \epsffile{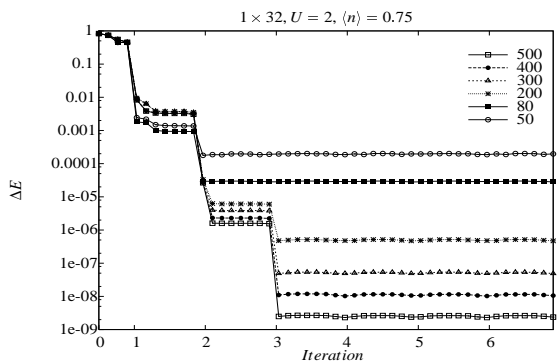}
\end{center}
\caption{Difference between the ground-state energy obtained from the {\it
finite-system}
DMRG with different number of iterations and number of states kept $m$, and the exact
energy calculated using Bethe Ansatz for a 32-site Hubbard model ($U=2$ and filling
$n=3/4$). Reprinted from \cite{whitebook} with permission.}
\label{figsisfinito}
\end{figure}

The method described above is usually called the \textit{infinite-system
algorithm} since the system size increases at each iteration. There is a way
to increase precision at each length $N$ called the \textit{finite-system
algorithm}. It consists of fixing the lattice size and zipping to and fro a couple of
times until convergence is reached. In this case and for the block
configuration [\textbf{B a a' B' }], $N=l+1+1+l^{\prime}$ where $l$ and $%
l^{\prime}$ are the number of sites in $B$ and $B^{\prime}$ respectively. In
this step the density matrix is used to project onto the left $l+1$ sites (see Fig. 
\ref{fig3}).
In order to keep $N$ fixed, in the next block configuration, the right block
$B^{\prime}$ should be defined in $l-1$ sites such that $N=(l+1)+1+1+(l-1)^{%
\prime}$. The operators in this smaller block should be kept from previous
iterations (in some cases from the iterations for the system size with $N-2$)
\cite{book}. With this method, a higher precision can be achieved (see Fig.
\ref{figsisfinito}). It is only within this finite-system algorithm that the total errors
are controlled (and are proportional) to the discarded error of the density matrix.

As an example of a general DMRG building step let's look at the one-dimensional
Heisenberg model
\begin{equation}
H=\sum_{i} {\bf S}_i {\bf
S}_{i+1}=S^z_iS^z_{i+1}+\frac{1}{2}(S^+_iS^-_{i+1}+S^-_iS^+_{i+1}).
\end{equation}
Suppose we want to connect two blocks [\textbf{$B_1$=B a }], defined by states $i,j$, with
[\textbf{$B_2$=a'B'}], defined by states $i',j'$,  to
form the Hamiltonian in the superblock, which we will diagonalize to get the energies and
eigenstates. Performing an external product between the operators we have stored from
previous iterations we have:


\begin{eqnarray}
[ H_{B_1 B_2}] _{ij;i'j'}
=&[ H_{B_1}] _{ii'}\delta _{jj'}+
[ H_{B_2}] _{jj'}\delta _{ii'} +
[ S^z_a] _{ii'}[ S^z_{a'}] _{jj'}
\label{eqnblockedheis} \\
&+\frac{1}{2}
[S_a^+] _{ii'}[ S_{a'}^-] _{jj'}
+\frac{1}{2}[ S_a^-] _{ii'}[ S_{a'}^+]
_{jj'}\nonumber
\end{eqnarray}

If fermionic models are to be considered, more operators should be kept, such
as, for example, the creation and destruction operators at the borders of the
blocks to perform the interblock hopping.

The calculation of static properties like correlation functions is easily
done by keeping the operators at each step and performing the corresponding
basis change and reduction, in a similar manner as done with the Hamiltonian
in each block\cite{white2}. The energy and measurements are calculated in
the superblock.

\subsection{The density matrix projection}

The density matrix leads naturally to the optimal states in the system as we
demonstrate below \cite{whitebook,scholl}.

Let
\begin{equation}
|\psi_0\rangle=\sum_{i,j=1}^M \psi_{ij}|i\rangle|j\rangle
\label{eq:5}
\end{equation}
be a state of the universe (system +
environment), having real coefficients for simplicity (here we will assume the same
configuration of system and environment so $M=M'$, but this assumption is not necessary). We want to obtain
a variational wave function $|\hat{\psi_0}\rangle$ defined in an optimally reduced space,
generated by the $m$ ``system" vectors $|\alpha\rangle=\sum_{i=1}^m u_{\alpha i} |i\rangle$
\begin{equation}
|\hat{\psi_0}\rangle=\sum_{\alpha=1}^m\sum_{j=1}^M a_{\alpha j}|\alpha\rangle|j\rangle
\end{equation}
such that the modulus of the difference with the true wave function \ref{eq:5},
\begin{equation}
||\psi_0\rangle-|\hat{\psi_0}\rangle|^2=1-2\sum_{\alpha i j}\psi_{ij}a_{\alpha j}u_{\alpha
i} + \sum_{\alpha j} a^2_{\alpha j}
\label{eq:min}
\end{equation}
is minimal w.r.t the $a_{\alpha j}$. Setting the derivative in these variables to zero, this condition leads to
\begin{equation}
\sum_{i}\psi_{ij}u_{\alpha i}=a_{\alpha j}.
\label{eq:alfa}
\end{equation}
If we define the density operator for a pure state of the ``universe" $\rho=|\psi_0\rangle\langle \psi_0|$,
the reduced density matrix of the ``system" is:
\begin{equation}
\rho_{ii'}=\sum_j\rho_{ij,i'j}=\sum_j \langle j|\langle i|\psi_0\rangle\langle 
\psi_0|i'\rangle|j\rangle=
 \sum_j\psi_{ij}\psi_{i'j},
\end{equation}
Replacing \ref{eq:alfa} in Eq. \ref{eq:min} and using the above expression 
we obtain the condition:
\begin{equation}
1-\sum_{\alpha i i'} u_{\alpha i} \rho_{ii'}u_{\alpha
i'}=1-\sum_{\alpha=1}^m\omega_{\alpha},
\end{equation}
where $u_{\alpha i'}$ change basis from $|i\rangle$ to $|\alpha\rangle$  and
$\omega_{\alpha}$ are the density-matrix eigenvalues.
The above expression is minimum when using the largest eigenvalues and corresponding
eigenvectors of the density matrix $\rho$, which are all positive or zero.
So we have shown that the density matrix leads naturally to the optimal reduced basis giving
the best approximation to the initial state. The last term corresponds to the discarded
error.

\subsubsection{Quantum information interpretation}

Another alternative and very interesting interpretation of the density matrix
stems from quantum information theory. The DMRG provides an exciting link
between strongly correlated systems and quantum information, by providing
another perspective into quantum phase transitions and interacting wave
functions. As stated by J. Preskill in \cite{preskill} ``The most challenging
and interesting problems in quantum dynamics involve understanding the
behaviour of strongly-coupled many-body systems[...]. Better ways of
characterizing the features of many particle entanglement may lead to new and
more effective methods for understanding the dynamical behaviour of complex
quantum systems"

In Ref.\cite{osborne} an interpretation of the correlation functions of systems
at criticality is given in terms of wave function entanglement.
Recent work\cite{verst2,verstraete,latorre} has
focused on the quantum information perspective of DMRG (see also \cite{galindo}).

To understand the entanglement between the two parts of a bipartite ``universe", say {\it
system} S and {\it environment} E  one can perform a Schmidt
decomposition \cite{nielsenchuang} of the wave function of the ``universe" (assuming it is
in a pure state), {\it i.e.}
writing the wave function as a product of states of S and E:

\begin{equation}
|\psi\rangle =\sum_{\alpha=1}^{min(N_S,N_E)} \sqrt{\omega_{\alpha}}|\alpha^S\rangle
|\alpha^E\rangle,
\end{equation}
where $N_S$ and $N_E$ are the Hilbert space sizes of system and environment respectively.
Tracing upon the environment we obtain the system density matrix

\begin{equation}
\rho_S=\sum_{\alpha=1}^{min(N_S,N_E)} \omega_\alpha |\alpha^S\rangle
\langle \alpha^S|.
\end{equation}

Similarly, tracing upon the system degrees of freedom, we obtain $\rho_E$.
Both density matrices have the same rank given by $r\le min(N_S,N_E)$. One can now affirm
that the state $|\psi\rangle$ is entangled if and only if the Schmidt rank $r>1$.

A quantitative measure of entanglement is given by the von Neumann entropy
\cite{bennet,legeza2003} defined in either subsystem,
(here for example in the subsystem S), by:
\begin{equation}
S_S=-Tr {\rho_S}\log_2 {\rho_S}=-\sum_\alpha \omega_\alpha^S \log_2 \omega_\alpha^S
\end{equation}

By keeping the highest eigenvalues $\omega_\alpha^S$ of the density matrix, for the general case, one obtains the
largest entropy $S$ and, hence, the maximum entanglement between system and environment.
It follows from the singular value decomposition theorem that, for a pure target state and any length of
system $S$ and environment $E$ blocks, $S_S+S_E+I=S_{universe}(=0$ for a pure target or universe state), where
$I$ is the mutual information of the blocks and measures the correlation between them. If both blocks are uncorrelated,
then $I=0$. It also stems from the above that for periodic boundary conditions, where the density matrix eigenvalues
$\omega_\alpha$ have a slower decay, the block entropy is also larger than for open boundaries. Therefore,
the DMRG will perform better for models or setups which have lower block entropy.

This quantum information perspective leads to an interesting analysis of the performance of
the DMRG in different dimensions. The entropy $S$ is proportional to the number of states to
be kept
in order to maintain the relevant information and it depends on system size $N$ ($S_{N}$).
Using geometric arguments in a (d+1)-dimensional field theory including a
(d-1)-dimensional hypersurface dividing the universe in two (S+E), it is shown that
the entropy, which
resides essentially at the surface, scales as the
hypersurface area \cite{callan}
\begin{equation}
S_N\;\alpha\; (N/\lambda)^{d-1}
\label{eq:14}
\end{equation}
were $\lambda$ is an ultraviolet cutoff. For one dimension ($d=1$) a more detailed
calculation leads to a logarithmic scaling for gapless critical systems having the universal form
$S_{N}=c/3 \ln(N) + \lambda_1$ (where $c$ is the central charge of the underlying conformal field theory 
and $\lambda_1$ is related to the ultraviolet cutoff). One also obtains a saturated entropy for
non critical, gapped systems\cite{latorre,vidal2002,korepin} when the system size exceeds the correlation length
(which is infinite for critical systems).
This shows that for two dimensions, as the number of relevant states to be kept increases
with system size, the DMRG isn't an appropriate method as it is conceived. It is, however,
quite reliable for sufficiently small systems. 

The influence of open boundary conditions on the entropy
of entanglement for critical XYZ spin chains was analyzed in \cite{laflorencie}, finding an additional alternating
term connected with the antiferromagnetic nature of the model.

There have been several recent publications exploiting the quantum information perspective of the DMRG.
For example, the time dependence
of the the block entropy in spin models with sudden changes in the anisotropies was studied in \cite{dechiara}.
To measure the entanglement between two halves of an anisotropic Heisenberg chain separated by an impurity bond,
the dependence of the block entropies (calculated using the density matrix as explained above) with system size and
anisotropy was done in \cite{zhao}, finding different behaviours between the antiferro and ferromagnetic cases.
The entropy of two sites in a one-dimensional system $S=-Tr {\rho_{i,i+1}}\log_2 {\rho_{i,i+1}}$
was used as a detector of phase transitions in spin and fermionic systems 
and applied to models like the bilinear-biquadratic $S=1$ Heisenberg chain,
the ionic Hubbard model and the neutral-ionic transition in a donor-acceptor 
model for molecular chains\cite{legezatwosite}.
A similar quantity was used in \cite{rissler} to calculate the entanglement between molecular orbitals. This two-site method
turns out to work better than the single-site entropy for the detection of quantum phase transitions (QPT) proposed in 
\cite{zanardi} since it comprises non local correlations and can detect phases with off-diagonal long range order.
An even better approach seems to be the detection of QPT by using block entropies as proposed in \cite{deng} (without
resorting to the DMRG) as also shown in \cite{zanardi}. 

\subsection{General remarks}

With respect to accuracy and convergence, the DMRG behaves in different ways depending on
the nature of the problem. Its success strongly relies on the existence of the so-called
{\it matrix product states} MPS, a simple version of which is the AKLT
(Affleck-Kennedy-Lieb-Tasaki) state for spin 1 chains \cite{AKLT}. The DMRG can be viewed as
a variational approach within this formulation \cite{ostlund,dukelskyMPS,verstraete}.
For cases where the quantum ground state can be represented exactly as a
matrix-product state, \"Ostlund and Rommer showed that the renormalization eventually
converges to a fixed point and the density matrix has a finite spectrum of
non vanishing eigenvalues (see also \cite{peschel1}, where a nice example is given
based on the non-hermitean $q$-symmetric Heisenberg model using corner transfer matrices).
Gapful models also lead to an excellent performance of the method, since the density matrix
eigenvalue spectrum decays exponentially and so does the truncated weight with increasing
the number of states kept $m$ \cite{peschelint,okunishi1}.
Instead, in the case of critical 1D models with algebraically decaying correlation
functions,
it has been shown that the eigenvalue spectrum decays much slower \cite{ors} which slows
down further with system size \cite{chungpeschel}. An analysis of how to circumvent the
problems arising near critical points by exploiting conformal field theory predictions using
multi-targeted sates is done in \cite{boschi} and, in \cite{verstcirac}, a mathematical
description of how matrix-product states approximate the exact ground state is given.
For two-dimensional systems it can be
shown that the eigenvalue decay of the density matrix is extremely slow, thus the
method becomes very unreliable for dimensions higher than one \cite{chungpeschel2}.
An interesting optimization study in momentum space using concepts of quantum information
entropy to select the best state ordering configurations was performed in Ref.
\cite{legeza2003} and also applied to quantum chemistry calculations. They conclude that the
optimum results
are obtained when the states with maximum entropy are located at the center of the chain. However,
a subsequent study by the same authors led to the conclusion that a criterion based on the entropy profile
alone does not lead to optimal ordering\cite{legeza2004} and further considerations are given based on
quantum data compression concepts.
In Ref.
\cite{verstraete} the DMRG's low performance with periodic boundary conditions is analyzed
and an alternative proposal based on highly entangled states is presented which leads to a much
higher precision, comparable to the open boundary conditions results. In a recent work\cite{whitesinglesite},
S. White proposes an interesting correction to the DMRG method which consists of including only a single site
added to the system, where the subsequent incompleteness of the environment block is compensated by considering
a corrected density matrix which takes into account fluctuations of the added site. Comparing this method to the 
standard two-added-site method he obtains a similar accuracy for open boundary conditions (but faster by a factor of 3)
and a much higher accuracy for periodic boundary conditions.

An analytical formulation combining the block renormalization group with
variational and Fokker-Planck methods is detailed in \cite{delgadorg}.
The connection of
the method with quantum groups and conformal field theory is treated in \cite
{sierraqg} and an interesting derivation of the reduced density matrix for integrable
fermionic and bosonic lattice models from correlation
functions was done in \cite{peschelcorr}.
The articles mentioned above give a deep insight into the essence of the DMRG method.

\subsection{Symmetries}

It is crucial to include symmetry in the DMRG algorithm since important reductions
in the Hilbert space size can be achieved by fixing the quantum numbers associated with
each symmetry. It
allows to consider a smaller number of states, enhance precision and obtain
eigenstates with definite quantum numbers.

For example, most of the models treated with DMRG conserve the total $z$ spin projection
$S_z$ and particle number in the whole system, so special care has to be taken when
constructing the superblock Hamiltonian, remembering that each block is constituted by
several states with different quantum numbers. It is easy to see that when the superblock
quantum number can be obtained as an addition of quantum numbers of each constituent part of the system,
the density matrix is block diagonal.

Non abelian or non additive symmetries like SU(2) or total spin are more
difficult to implement. Sierra and Nishino showed that Interaction-round-a-face
(IRF) Hamiltonians represent Hamiltonians with a continuous symmetry and
developed a DMRG algorithm appropriate to this model\cite{sierrairf}. This
method was applied to several spin models with excellent performance
\cite{sierrairf,wada}. Total spin conservation, continuous symmetries and
parity have been treated also in \cite{ian,ramaseshasim,ducroo,affleck,simetrias} and
recently used in Ref. \cite{carbon} in carbon nanotubes, where flat band
ferromagnetism was found.

Contrary to momentum-space DMRG (see Sect 5.1), in real space the total momentum cannot be fixed, because
there is no way to fix the corresponding phases in the states that diagonalize the density
matrix. However, a very good approximation to having states with fixed momenta can be done by
keeping the appropriate target states \cite{karendin}.

\section{Applications}

Since its development in 1992, the number of papers using DMRG has grown enormously
and other improvements to the method have been performed. For example, since
1998 there have been around 80 papers a year using DMRG. There have also
been several improvements to the method and it is now used in areas that are
very different to the original strongly correlated electron system field. We
would like to mention some applications where this method has proved to be
useful. Other applications related to further developments of the DMRG will
be mentioned in subsequent sections.


An important benchmark in calculations using DMRG was achieved by White
and Huse \cite{white3} when calculating the spin gap in a $S=1$ Heisenberg
chain obtaining $\Delta=0.41050 J$ with unprecedented accuracy. They also calculated very accurate spin
correlation functions and excitation energies for one and several magnon
states and performed a very detailed analysis of the excitations for
different momenta. They obtained a spin correlation length of 6.03 lattice
spacings. Simultaneously S{\o}rensen and Affleck\cite{sorensen} also
calculated the structure factor and spin gap for this system up to length
100 with very high accuracy, comparing their results with the nonlinear $%
\sigma$ model. In a subsequent paper\cite{sorensen2} they applied the DMRG to
the anisotropic $S=1$ chain, obtaining precise values for the Haldane gap. They
also performed a detailed study of the $S=1/2$ end excitations in an open
chain. Thermodynamical properties in open $S=1$ chains such as specific heat,
electron paramagnetic resonance (EPR) and magnetic susceptibility calculated
using DMRG gave an excellent fit to experimental data, confirming the existence
of free spins 1/2 at the boundaries\cite{batista,batista2,batista3,batista4}. A 
related problem,
\textit{i.e.} the effect of non-magnetic impurities in spin systems (dimerized,
ladders and 2D) was studied in \cite{laukamp,ng}. In addition, the study of
magnon interactions and magnetization of $S=1$ chains was done in
\cite{affleck3}, supersymmetric spin chains modelling plateau transitions in
the integer quantum Hall effect in \cite{marston4} and ESR studies in these
systems was considered in \cite{sieling}. For larger integer spins there have
also been some studies. Nishiyama and coworkers\cite{nishi} calculated the low
energy spectrum and correlation functions of the $S=2$ antiferromagnetic
Heisenberg open chain. They found $S=1$ end excitations (in agreement with the
Valence Bond Theory). Edge excitations for other values of $S$ have been
studied in Ref. \cite{qinedge}. Almost at the same time Schollw{\"o}ck and
Jolicoeur\cite{uli1} calculated the spin gap in the same system, up to 350
sites, ($\Delta=0.085 J$), correlation functions that showed topological order
and obtained a spin correlation length of 49 lattice spacings. More recent accurate
studies of $S=2$ chains are found in \cite {wangqin,wada,capone} and of $S=1$
chains in staggered magnetic fields \cite {yulu2} including a detailed
comparison to the non-linear sigma model in \cite{ercolesi}. In Ref. \cite{qin}
the dispersion of the single magnon band and other properties of the $S=2$
antiferromagnetic Heisenberg chains were calculated and the phase diagram
of S=1 bosons in 1D lattices, relevant to recent experiments in optical lattices
was studied in \cite{rizzi}.

Concerning $S=1/2$ systems, DMRG has been crucial for obtaining the
logarithmic corrections to the $1/r$ dependence of the spin-spin correlation
functions in the isotropic Heisenberg model \cite{karen12}. For this, very
accurate values for the energy and correlation functions were needed. For $%
N=100$ sites an error of $10^{-5}$ was achieved keeping $m=150$ states per
block, comparing with the exact finite-size Bethe Ansatz results. For this
model it was found that the data for the correlation function has a very
accurate scaling behaviour and advantage of this was taken to obtain the
logarithmic corrections in the thermodynamic limit. Other calculations of
the spin correlations have been performed for the isotropic \cite
{boos,shiroishi} and anisotropic cases \cite{hiki}. Luttinger liquid
behaviour with magnetic fields have been studied in \cite{luttinger},
field-induced gaps and string order parameters in \cite{lou2} and \cite{lou3}
respectively, anisotropic systems in \cite
{caprara,hieida} and the Heisenberg model with a weak link in \cite{byrnes}.
An analysis of quantum critical points and critical behaviour in spin chains
by combining DMRG with finite-size scaling was done in \cite{marston2}. Spin systems in more
complex lattices like the Bethe lattice were considered in \cite{pastor} and the effect of
twisted boundary magnetic fields in \cite{zhuo}.

Similar calculations have been performed for the $S=3/2$ Heisenberg chain
\cite{karen32}. In this case a stronger logarithmic correction to the spin
correlation function was found. For this model there was interest in
obtaining the central charge $c$ to elucidate whether this model corresponds
to the same universality class as the $S=1/2$ case, where the central charge
can be obtained from the finite-size scaling of the energy. Although there
have been previous attempts\cite{adriana}, these calculations presented
difficulties since they involved also a term $\sim 1/\ln^3N$. With the DMRG
the value $c=1$ was clearly obtained.

In Ref. \cite{yamashita}, DMRG was applied to an effective spin Hamiltonian
obtained from an SU(4) spin-orbit critical state in 1D. Other applications
were done to enlarged symmetry cases with SU(4) symmetry in order to study
coherence in arrays of quantum dots\cite{onu}, to obtain the phase diagram
for 1D spin orbital models\cite{affleck2} and dynamical properties in a
magnetic field\cite{haas2}.

Dimerization and frustration have been considered in Refs. \cite
{bursill,dim1,dim2,dim3,dim4,dim5,dim6,kaburagi,hikihara2} and alternating
spin chains in \cite{patialt}.

Several coupled spin chains (ladder models) have been
investigated in \cite{noackchain,kawaguchi,trumper, 
roger,hikihara3,capriotti,rodriguezlaguna}, spin
ladders with cyclic four-spin exchanges in \cite{schmid,hikihara,
nunner,honda} and Kagome antiferromagnets in \cite{patising}. Zigzag spin
chains have been considered in \cite{maeshima,itoi, yulu}, spin chains of
coupled triangles in \cite{triang,schotte,gendiar3}, triangular Ising models in \cite{otsuka}
and three-leg spin tubes in \cite{fouet}.
As the DMRG's
performance is optimal in open systems, an interesting analysis of the
boundary effect on correlation functions is done in \cite{scalapino}.
Magnetization properties and plateaus for quantum spin chains and ladders \cite
{tandon,lou,wanglu,silva,gu} have also been studied. An interesting review on the
applications to some exact and analytical techniques for quantum magnetism
in low dimension, including DMRG, is presented in \cite{patisen}.

There has been a great amount of applications to fermionic systems such as
1D Hubbard, ionic Hubbard and t-J models \cite{noackhubb,penc,bulut,malvezzi,eric,zhang2,arno,
baeriswyl,aoki,daul2,qin2,armando,weisse}, Luttinger liquids with boundaries\cite
{kurt}, the Falicov-Kimball model \cite{falicov}, the quasiperiodic
Aubry-Andre chain\cite{schuster} and Fibonacci-Hubbard models\cite{fibonacci}.
A recent calculation obtains accurate values of the Luttinger-liquid parameter $K_{\rho}$
in Hubbard and spin chains from static correlation functions \cite{ejima}.
It has also been applied to field theory\cite{gaite,phi4}. The method has
been very successful for several band Hubbard models\cite{srinivasan,xavier2} and extended Hubbard models
modelling chains of $CuO_4$ plaquettes\cite{hozoi},
Hubbard ladders \cite{hubbardladder,vojta,liang2,ramasesha3,edegger2} and t-J ladders\cite
{scalapino2}. Recent calculations in doped Hubbard ladders include the observation of stripes
in the 6-leg ladder for large values of the interaction parameter and not for low values \cite{hager},
the study of orbital currents and charge density waves in ladders of up to 200 rungs \cite{fjaerestad} and
existence of charge order induced by electron-lattice interaction in coupled ladders \cite{edegger}.
Ring exchange on doped two-leg ladders were investigated in \cite{roux}.
Also several coupled chains at different dopings have been
considered \cite{fermion,nishimoto} as well as flux phases in these systems
\cite{marston}.  Time reversal symmetry-broken fermionic ladders have been
studied in \cite{time} and power laws in spinless-fermion ladders in \cite
{bourbonnais}. Long-range Coulomb interactions in the 1D electron gas and
the formation of a Wigner crystal was studied in \cite{ziosi}. Several
phases including the Wigner crystal, incompressible and compressible liquid
states, stripe and pairing phases, have been found using DMRG for 2D
electrons in high magnetic fields considering different Landau levels\cite
{landau,landau1,landau2}. Persistent currents in mesoscopic systems have been 
considered in
\cite{rings}.

Impurity problems have been studied for example in one- \cite{teo} and
two-impurity \cite{egger} Kondo systems, in spin chains \cite{impchains} and
in Luttinger Liquids and Hubbard models\cite{meden,andergassen}. There have also been applications to
Kondo and Anderson lattices \cite
{kondoins,kondolatt,kondoneck,kondoxavier,kondonecklace,kondojuoza,
kondoguerrero,watanabe,ian,xavier,masa}, Kondo lattices with localized $f^{2}$
configurations \cite{wata}, the two-channel Kondo lattice on a ladder \cite
{kondomoreno} and on a chain \cite{schauerte}, a t-J chain coupled to localized Kondo spins\cite{chen} and
ferromagnetic Kondo models for manganites \cite{riera, kondodaniel,imada,neuber}.
See also Refs. \cite{daniel,mcculloch} for recent results for the antiferro and
ferromagnetic Kondo lattice models.

The DMRG has also been generalized to 1D random and disordered systems, and
applied to the random antiferromagnetic and ferromagnetic Heisenberg chains
\cite{hida,lajko}, including quasiperiodic exchange modulation\cite{hida1} and a
detailed study of the Haldane\cite{hida2} and Griffiths phase\cite{griffiths}
in these systems. Strongly disordered spin ladders have been considered in
\cite{spinladders}. It has also been used in disordered Fermi systems such
as the spinless model\cite{schmitt,schmitt2}. In particular, the transition
from the Fermi glass to the Mott insulator and the strong enhancement of
persistent currents in the transition was studied in correlated
one-dimensional disordered rings\cite{jala}. Disorder-induced crossover
effects at quantum critical points were studied in \cite{carlonigloi}.

There have been recent applications of the DMRG in nanoscopic devices such as
transport properties through quantum dots \cite{maruyama}, the study of the
influence of interactions in a reservoir on the levels of a quantum dot coupled
to it \cite{sade} and charge sensing in quantum dots \cite{berkovits}. A very
interesting application of the DMRG to study single wall carbon nanotubes,
where the tube is mapped onto a 1D chain with longer range interactions
(depending on the quirality) can be found in Ref. \cite{ye} and the study of edge states
in doped nanostructures in \cite{chernyshev}. Inspired in recent
experiments of trapped bosonic atoms \cite{paredes}, a one dimensional Bose gas
was studied using ab-initio stochastic simulations at finite temperatures
covering the whole range from weak to strong interactions\cite{schmidt}. They used a
discretized version of the model and a block factorization of the kinetic
energy, using the density matrix to select the most relevant states from each
block. Also models for trapped ions have been studied using DMRG, where different 
quantum phases were obtained\cite{deng2}. A recent study concerns the colossal electroresistance
(due to band bending) through the interface between a metal and a strongly interacting system 
(in 1D) in the presence of an electric potential obeying the Poisson equation. In this case the use 
of the finite-size method played an important role in the convergence of the ground state due to
the local density-dependent potential.\cite{oka1}

\section{Bosonic degrees of freedom}

A vast research area in correlated systems concerns bosonic degrees of freedom.
For example, phonons play fundamental role in models of currently interesting
correlated systems such as high-Tc superconductors, manganites,
organic compounds, or nanoscopic systems. 

A significant limitation to the DMRG method is that it requires a finite
basis and calculations in problems with infinite degrees of freedom per site
require a large truncation of the basis states\cite{moukouri2}. However,
Jeckelmann and White developed a way of including phonons in DMRG
calculations by transforming each boson site into several artificial
interacting two-state pseudo-sites and then applying DMRG to this
interacting system\cite{jeck} (called the ``pseudo-site system"). The proposal
is based on the fact that DMRG is much better able to handle several
few-states sites than few many-state sites\cite{noackberlin}. The key idea
is to substitute each boson site with $2^N$ states into $N$ pseudo-sites
with 2 states\cite{jeckbook}. They applied this method to the Holstein model
for several hundred sites (keeping more than a hundred states per phonon
mode) obtaining negligible error. In addition, to date, this method is
the most accurate one to determine the ground state energy of the polaron
problem (Holstein model with a single electron). For a recent comprehensive review
on exact numerical methods in electron-phonon problems see \cite{jeckreview}.

This method has been applied recently to the calculation of 
pairing, CDW and SDW correlations in the Holstein-Hubbard model with competing electron-electron and 
electron-phonon interactions, finding an enhanced superconducting exponent when next-nearest neighbour
hopping is added\cite{tezuka}. They propose here a way to improve the bare repulsion felt
by the electrons in the newly added pseudo-sites in the infinite-size algorithm
by adding a chemical potential to these sites, finding lower ground state energies. 
An interesting result was obtained in \cite{matsuedaholstein} using this method combined with the dynamical
DMRG (see Sect.\ref{sec:lanczos}) to calculate single-particle spectra in one-dimensional Mott insulators 
(the Holstein-Hubbard model at half filling in this case). They observe that the charge-spin separation
characteristic of 1D systems is robust against electron-phonon coupling.

\begin{figure}[tbp]
\begin{center}
\epsfxsize=3.0in \epsfysize=2.0in \epsffile{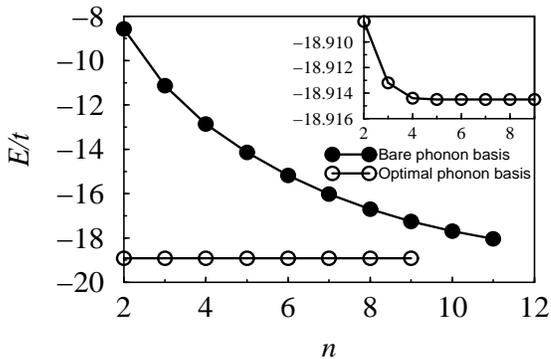}
\end{center}
\caption{Ground-state energy of a 4 site, half-filled Holstein system vs. the number 
of phonon states kept on each site of the lattice, for the optimal and bare phonon 
basis. Reprinted from \cite{jeck2} with permission.}
\label{figfon}
\end{figure}

An alternative method (the ``Optimal phonon basis")\cite{jeck2} is a
procedure for generating a controlled truncation of a large Hilbert space,
which allows the use of a very small optimal basis without significant loss
of accuracy. The system here consists of only one site and the environment
has several sites, both having electronic and phononic degrees of freedom.
The density matrix is used to trace out the degrees of freedom of the
environment and extract the most relevant states of the site in question. In
following steps, more bare phonons are included to the optimal basis
obtained in this way. This method was successfully applied to study the Holstein model
(Fig.\ref{figfon}), the
interactions induced by quantum fluctuations in quantum strings as an
application to cuprate stripes\cite{nishiyama3}, the dissipative
two-state system \cite{nishiyama5} and, recently, the spin-Peierls model \cite{barfordpeierls}.
A variant of this scheme is the ``four
block method", as described in \cite{bursillphonon}. They obtain very
accurately the Luttinger liquid-CDW insulator transition in the 1D Holstein
model for spinless fermions.

The method has also been applied to pure bosonic systems such as the
disordered bosonic Hubbard model\cite{krish}, where gaps, correlation
functions and superfluid density are obtained. The phase diagram for the
non-disordered Bose-Hubbard model, showing a reentrance of the superfluid
phase into the insulating phase was calculated in Ref. \cite{monien}. It has
also been used to study a chain of oscillators with optical phonon spectrum
\cite{chung} and optical phonons in the quarter-filled Hubbard model for
organic conductors\cite{maurel1}.

\section{Generalized DMRG}
Since its development, the DMRG has been widely improved in several directions in order to
deal with different kinds of problems. It has been generalized to treat models where a
representation in momentum or
energy space is more appropriate. Based on these ideas, an important extension towards
Quantum Chemistry and the calculation of physical properties of small grains and nuclei
was performed. Below we describe each of these ideas.

\subsection{Momentum representation}

In 1996, Xiang \cite{xiang2} proposed an alternative formulation of the DMRG in
momentum space, rather than in real space, which can be used also in two or more
dimensions, it deals better with periodic boundary conditions and has the advantage that
the total momentum quantum number can be fixed.
This algorithm has inspired the extension of
the DMRG to finite fermionic systems, small grains and nuclear physics.
The main idea is to write the Hamiltonian in momentum space. If the original model is
long ranged, as with Coulomb interactions, then it will turn local in the reciprocal
space and the DMRG could be used in the normal way, where now each ``site" is a state
with definite momentum. The most convenient growth procedure for real space short-ranged 
Hamiltonians is to start using
one-particle states lying near the Fermi energy, above and below, and subsequently add
additional more distant states. For longer ranger models other orderings are to be
considered \cite{legeza2003}.

In his original work, Xiang applied these ideas to the 1D and 2D Hubbard models with
nearest-neighbour interaction, defined in general as:

\begin{equation}
H=-t\sum_{<ij>\sigma }\left( c_{i\sigma }^{\dagger }c_{j\sigma
}+c_{j\sigma }^{\dagger }c_{i\sigma }\right)
+U\sum_{i}n_{i\uparrow }n_{i\downarrow } ~. \label{Hub}
\end{equation}
The operator $c^{\dagger}_{i \sigma}~(c_{i\sigma})$ creates
(annihilates) an electron on site $i$ with spin projection
$\sigma$, $n_{i \sigma}=c^{\dagger}_{i \sigma} c_{i \sigma}$,
$t$ is the hopping parameter and $U$ the on-site Coulomb interaction.

In momentum space it reads:

\begin{equation}
H = \sum_{\bf{k}\sigma} \epsilon_{\bf{k}} {n}_{\bf{k}\sigma}
~+~ \frac{U}{N} \sum_{\bf{p,k,q}} c^{\dagger}_{\bf{p-q}\uparrow}
c^{\dagger}_{\bf{k+q}\downarrow} c_{\bf{k}\downarrow}
c_{\bf{p}\uparrow} ~, \label{kHub}
\end{equation}
where
${n}_{\bf{k}\sigma} = c^{\dagger}_{{\bf k}\sigma}c_{{\bf
k}\sigma}$
is the number operator for particles with spin $\sigma$ and
momentum $\bf k$ and $\epsilon_{\bf{k}}$ is the energy dispersion of the particles.

By defining appropriate operator products using the original ones, the number
of operators to be kept can be greatly reduced and the convergence becomes
faster and more accurate. Other authors \cite{nishimoto3,legeza2003} have
improved the method and performed more systematic and interesting analysis on
the order of states to be kept, depending on the kind of interactions in the
model and very accurate dispersion relations have been obtained in
\cite{nishimoto}.
 The method has a better performance
than the real space DMRG in two dimensions for low values of the interaction (as seen
in 4x4 lattice calculations), the accuracy of the method becoming worse for larger
values of $U$.
Recently, an interesting algorithm for the calculation of eigenstates with definite momentum
has been put forward in \cite{porras}, based on the PEPS (projected entangled pair
states). In this case additional degrees of freedom are added for each site and the particular
momentum quantum number is fixed by constructing a ``projected entangled-multiparticle state"
(PEMS). It has been successfully applied to calculate excitation spectra in the bilinear-biquadratic
$S=1$ Heisenberg model.

\subsection{Grains, nuclei and high energy physics}

Another successful extension to the DMRG is the calculation of physical properties of
small fermionic systems, like superconducting grains and models for nuclei. This was
developed by Dukelsky, Sierra and collaborators \cite{duksierra}. For a review on
these applications see the recent publication \cite{dukelskyreview}.

A typical and general Hamiltonian for these finite systems consists of one- and
two-particle terms of the form

\begin{equation}
H=\sum_{ij}T_{ij}c_{i}^{\dagger
}c_{j}+\sum_{ijkl}V_{ijkl}c_{i}^{\dagger }c_{j}^{\dagger
}c_{l}c_{k} ~, \label{H1}
\end{equation}
where $T_{ij}$ represents the kinetic energy and $V_{ijkl}$ the residual
two-body interaction between the effective particles.

\begin{figure}
\begin{center}
\hspace{1.5cm}\epsfxsize=5.5cm \epsfysize=7.2cm\epsfbox{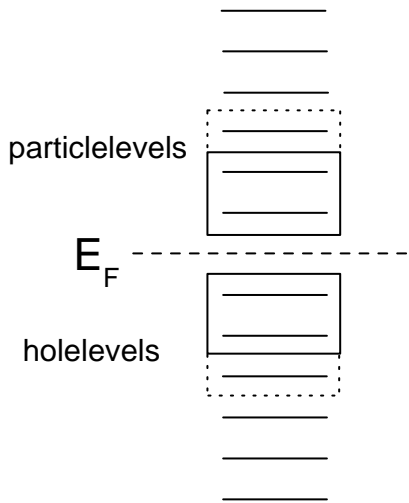}
\end{center}
\caption{ DMRG method for finite fermionic systems where the added "sites"
lie successively at larger distances from the Fermi energy.}
\label{figlevels}
\end{figure}

Starting from the single particle levels and a certain filling which
defines the Fermi
energy, one can define particle and hole blocks in a similar manner as the real space
DMRG (see Fig. \ref{figlevels}). The empty and occupied levels are the particle and hole levels
respectively (referring to the character of the single-particle excitations). 
By successively adding levels at increasing distance from the Fermi
level, the DMRG renormalization is applied. This might not be the appropriate level
ordering for DMRG for more complex Hamiltonians \cite{legeza2003,dukelskyreview}.

Within this scheme, superconductivity in small grains has been studied using
the half-filled reduced BCS Hamiltonian
\begin{equation}
H = \sum_{j=1}^{\Omega} \sum_{\sigma=\pm}(\epsilon_j -\mu)
c^\dagger_{j\sigma} \; c_{j\sigma} - \lambda d
\sum_{i,j=1}^{\Omega} \; c^\dagger_{i+} c^\dagger_{i-} c_{j-}
c_{j+} ~, \label{D1}
\end{equation}
where $i,j=1,2,\dots ,\Omega $ label the single-particle
energy levels, $c_{j\sigma }$ are the electron annihilation
operators associated with the two time-reversed states
$|j,\sigma=+\rangle$ and $|j,\sigma =-\rangle$ , $\mu $ is
the chemical potential, $\lambda $ is an adimensional BCS
coupling constant and $d$ is the single-particle energy level spacing (assumed to be constant).
By using DMRG a smooth superconducting-normal state crossover with decreasing
system size was found, a more reliable result than the abrupt transition obtained using
exact diagonalization and BCS approximation.
This method was also used for two superconducting grains coupled by tunneling
\cite{gobert}.

Other recent applications have been in nuclear shell model calculations where a two-level
pairing and pairing plus quadrupole interactions model have been
considered\cite{dukelsky}. Here the nucleus is modelled by
completely filled core shells and valence orbitals partially filled
with protons and neutrons where the starting states are the Hartree Fock functions.
A realistic calculation for the $^{24}$Mg nucleus showed very slow convergence mainly
due to the lack of improvement of the finite-system algorithm in this basis
and the fact that total angular momentum cannot be explicitly fixed.
A recent work obtained the {\it sd}- and {\it fp}-shell models for
$^{28}$Si and
$^{56}$Ni respectively in a different basis using proton-neutron wave function
factorization which allows fixing total angular momentum plus the finite-system
algorithm \cite{papenbrock}. The DMRG has also been extended to study complex-symmetric density
matrices which describe many-body open quantum systems (where resonant and non-resonant levels are considered)
and applied to the unbound nucleus ${^7}He$ with possible further applications to open quantum dots,
open microwave resonators, etc.\cite{rotureau}

A very interesting and successful application is a recent work in High
Energy Physics\cite{delgadoqcd}. Here the DMRG is used in an asymptotically
free model with bound states, a toy model for quantum chromodynamics, namely
the two dimensional delta-function potential. For this case an algorithm
similar to the momentum space DMRG\cite{xiang2} was used where the block and
environment consist of low and high energy states respectively. The results
obtained here are much more accurate than with the similarity
renormalization group\cite{wilson2} and a generalization to
field-theoretical models is proposed based on the discreet light-cone
quantization in momentum space\cite{dlcq}.

\subsection{Molecules and Quantum Chemistry}

Using the standard real space DMRG, there have been several applications to molecules and polymers, such as the
Pariser-Parr-Pople (PPP) Hamiltonian for a cyclic polyene\cite{ppp,pppreview} (where
long-range interactions are included), magnetic Keplerate molecules \cite
{keplerate}, molecular Iron rings\cite{bruce}, spin crossover molecules \cite{timm}
  and polyacenes considering
long range interactions \cite{polyacenes}. The application to
conjugated organic polymers was performed by adapting the DMRG to take into
account the most important symmetries in order to obtain the desired excited
states\cite{ramaseshasim}. Also conjugated one-dimensional semiconductors \cite
{barford} have been studied, in which the standard approach can be extended
to complex 1D oligomers where the fundamental repeat is not just one or two
atoms, but a complex molecular building block. Relatively new fields of
application are the calculation of dynamical properties in the
Rubinstein-Duke model for reptons \cite{reptons} and excitons in dendrimer
molecules\cite{dendrimers}.

\begin{figure}
\begin{center}
\epsfxsize=4.0in \epsfysize=2.0in \epsffile{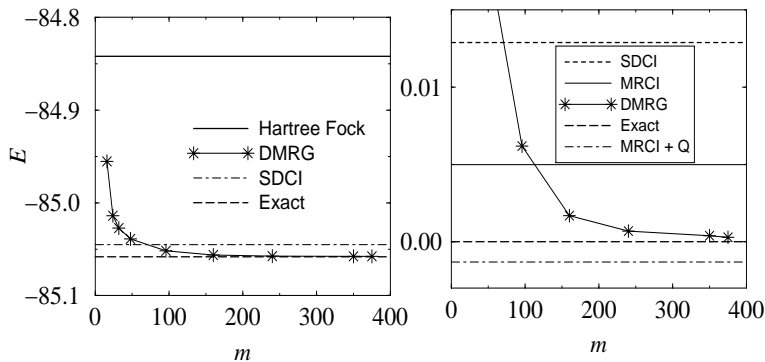}
\end{center}
\caption{Ground state energy (left) and error (right) of a water molecule in a 25 orbital
basis using several methods. The DMRG results are plotted as a funtion of the number of
states kept, $m$ (not relevant for the other energies). Reprinted from \cite{whitemol}
with permission.}
\label{figmol}
\end{figure}

Another very important problem is the \textit{ab initio} calculation of
electronic states in Quantum Chemistry.
The standard approaches consist of the Density Functional Theory and the Hartree Fock starting point
to calculate complex electronic structure (using for example the Configuration Interaction approach).
When electron interactions are included, relatively small clusters can be considered
due to the large Hilbert spaces involved.
To be able to treat larger molecules, in 1999, White and Martin \cite{whitemol} proposed a way of using the DMRG
for quantum chemistry calculations, based on the momentum representation
DMRG \cite{xiang2}. They considered a general model for molecules, Eq (\ref{H1}), where the
electron-nucleus interaction in considered in the first term and the operators indices represent
orbitals in the Hartree Fock basis (see also \cite{whiteorth,
legezaqc,daulwhite}).
Here, DMRG is applied within the conventional quantum
chemical framework of a finite basis set with non-orthogonal basis functions
centered on each atom. After the standard Hartree-Fock (HF) calculation in
which a Hamiltonian is produced within the orthogonal HF basis, DMRG is used
to include correlations beyond HF, where each orbital is treated as a
``site" in a 1D lattice. One important difference with standard DMRG is
that, as the interactions are long-ranged, several operators must be kept,
making the calculation somewhat cumbersome. However, very accurate results
have been obtained in a check performed in a water molecule (keeping up to
25 orbitals and $m\simeq 200$ states per block), obtaining an offset of
$2.4 10^{-4}$Hartrees with respect to the exact ground state energy\cite{bau}, a
better performance than any other approximate method\cite{whitemol} (see
Fig.\ref{figmol}).

In order to avoid the non-locality introduced in the treatment explained
above, White introduced the concept of \textit{orthlets}, local, orthogonal
and compact wave functions that allow prior knowledge about singularities to
be incorporated into the basis and an adequate resolution for the cores\cite
{whiteorth}. The most relevant functions in this basis are chosen via the
density matrix. An application based on the combination with the momentum
version of DMRG is used in \cite{legezaqc} to calculate the ground state of
several molecules.
In Ref. \cite{mitrushenko}, the efficiency of the DMRG method was analyzed. Although it works very well
for some molecules, like Be$_2$, HF, H$_2$O, CH$_4$, calculations in the N$_2$ molecule  were not satisfactory.
The reasons for the failure could be related to the level ordering taken for the DMRG algorithm, although it
is not conclusive. A very careful analysis of the method presented in a pedagogical way can be found in Ref.
\cite{chan}.

In a recent work, quantum information theory concepts were used together with DMRG to calculate orbital interaction in
different molecules, such as LiF, CO, $N_2$ and $F_2$ \cite{rissler}. Here the interaction between electrons in two orbitals
in a many-body wave function was calculated 
as the difference between the entanglement of both orbitals taken together and individually (always belonging to the 
same DMRG ``system") for a given wave function. An interesting analysis of the relevance of the level ordering within the DMRG
growing procedure is given.

\section{Higher dimensions ($D>1$)}

\begin{figure}[tbp]
\vspace{1cm}
\begin{center}
\epsfxsize=3.0in \epsfysize=2.0in \epsffile{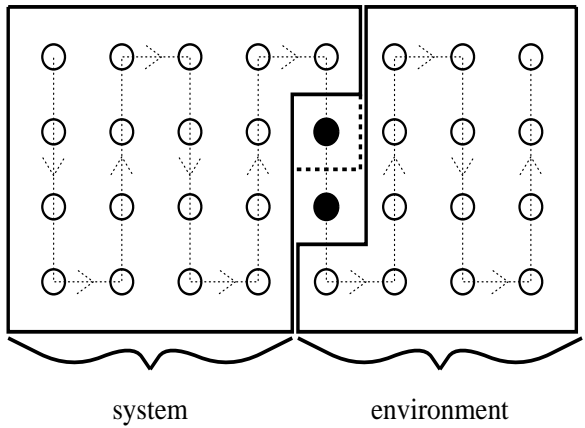}
\end{center}
\caption{General sweeping and superblock configuration for a two-dimensional lattice
(from \cite{whitebook}, with permission)}
\label{fig2d}
\end{figure}

The possibility of using DMRG in two dimensions is currently one of the most
challenging problems. As we mentioned above equation \ref{eq:14}), due to the 
increasing number of
states to be kept in dimensions higher than one, the DMRG is not suited for
these cases. By analyzing analytically the spectra of a two-dimensional system
of coupled oscillators, Chung and Peschel \cite{chungpeschel2} showed that the
density matrix eigenvalues decay much slower than in the one-dimensional case.
 However, if the system size is kept
relatively small, in two dimensions one can reach larger systems than with exact
diagonalization methods and the zero temperature results are more reliable than with Monte
Carlo methods.
The first attempts to extend the DMRG to two dimensions considered different
one-dimensional kind of paths like the one shown in figure \ref{fig2d}, where the 
DMRG is
used in the standard way.
The main drawback of this method is that many operators have to be kept and the interactions
are long-ranged in this 1D mapping.

Quite large quasi-2D systems can be reached, for example in \cite{liangpang}
where a 4x20 lattice was considered to study ferromagnetism in the
infinite-U Hubbard model; the ground state of a 4-leg t-J ladder in \cite{w1}%
; the one- and two-hole ground states in 9x9 and 10x7 t-J lattices in \cite{w2}%
; a doped 3-leg t-J ladder in \cite{w3}; the study of striped phases in \cite
{arrigoni}; domain walls in 19x8 t-J systems in \cite{w4}; the 2D t-J model
in \cite{bishop} and the magnetic polaron in a 9x9 t-J lattice in \cite
{white5}. Also big $CaV_4O_9$ spin-1/2 lattices reaching 24x11 sites\cite
{cavo} have been studied as well as a recent study of the spin liquid phase in the 
anisotropic triangular Heisenberg model in up to 8x18 lattices \cite{weng}.

There have also  been some recent alternative attempts to implement
DMRG in two and higher dimensions with clever block arrangements \cite
{su,nishimoto3,nishino5,sierra3d,henelius,farnell} but the performances are still
poorer than in 1D. A recent extension using a novel DMRG algorithm for
highly anisotropic spin systems has shown promising results\cite{moukouri5}. Here 
a two-step method is proposed which is based on the diagonalization of a single chain
to obtain the lowest-lying states and then using DMRG to couple these chains in a 1D manner.
As long as the energy width of the retained states is much larger than the interchain coupling,
this method yiels results which are comparable to Quantum Monte Carlo calculations.

\begin{figure}[htbp]
\begin{center}
\epsfxsize=3.9in \epsfysize=2.4in \epsffile{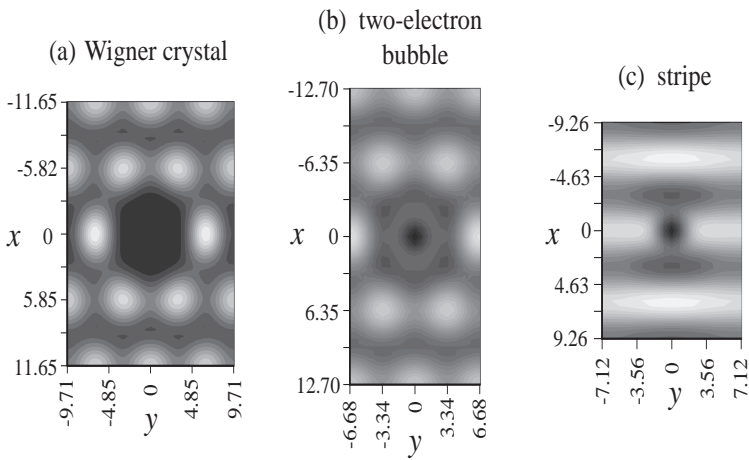}
\end{center}
\caption{Pair correlation function of a) a striped state, b) a two-electron bubble state and
c) a Wigner crystal
(from \cite{shibatareview}, with permission)}
\label{figlandau}
\end{figure}

Another problem in physics that has been handled using DMRG is that of electrons in
a high magnetic field. Shibata and Yoshioka \cite{landau,landau1,landau2} used the 
eigenstates of free electrons
in a perpendicular magnetic field in the Landau gauge to represent the local orbitals.  
By choosing these
single particle states as effective sites, where the wave functions are parametrized 
by only two quantum numbers
(the Landau level and the $x$ component of the center of coordinates of an electron 
in a cyclotron motion),
 it provides a natural mapping onto a 1D system (see also
\cite{shibatareview}). The long-range Coulomb interactions between the electrons stabilizes various different electronic
states depending on the filling of the Landau levels. If the magnetic field is larger than the characteristic
Coulomb interactions, the electrons in filled Landau levels can be considered inert and the ground state is
determined by the partially filled level.
 The ground state properties for different filling factors were characterized using
the expectation value of the pair correlation function
\begin{equation}
g\left( \mathbf{r}\right) \equiv \frac{L_{x}L_{y}}{N_{e}\left(
N_{e}-1\right) }\left\langle \Psi \right\vert \sum_{i\neq j}\delta
\left( \mathbf{r}+\mathbf{R}_{i}-\mathbf{R}_{j}\right) \left\vert
\Psi \right\rangle ~,
\end{equation}
where $\mathbf{R}_{i}$ is the center coordinate of the
$i$th electron's wave function, $\mathbf{r}$ is the relative distance between the
pair of electrons and $L_{x,y}$ are the unit cell dimensions.
In \cite{landau} the phase diagram for the third Landau level 
was obtained, showing a Wigner crystal for low filling
factors, up to $\nu =0.24$, where a 2-electron bubble phase forms. For $\nu > 0.38$, a
striped phase shows up (see Fig. \ref{figlandau}) and in \cite{landau2} an analysis
of the excitonic behaviour of the $\nu=1$ bilayer quantum Hall systems was done.

\begin{figure}
  \centering
  \epsfxsize=3.0in \epsfysize=2.0in \epsffile{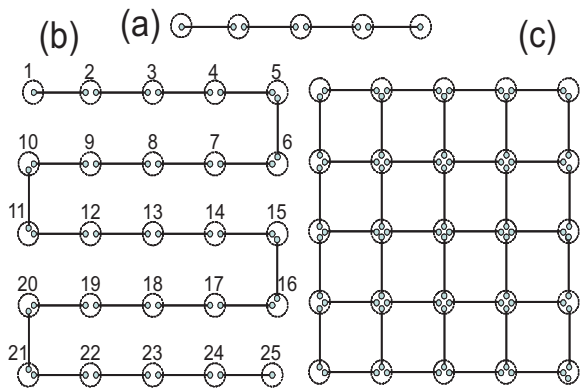}
  \caption{Schematic representation of MPS (a) 1 dimension
 and (b) 2 dimensions. (c) representation of 2D PEPS. The bonds represent
  pairs of maximally entangled D--dimensional auxiliary spins and the
  circles denote projectors that map the auxiliary spins
  to the physical ones. Reproduced from \cite{verst2} with permission.}
  \label{figverst}
\end{figure}

A very promising approach towards two dimensional calculations was
recently  proposed by Verstraete and Cirac \cite{verst2}, based on the {\it
projected entangled-pair-states} (PEPS). It relies on the matrix-product states
(MPS) approach to describe the variational wave functions which is the basis of
success of the DMRG. Following Ref. \cite{verst2}'s notation, let's represent
the 1D state $|\Psi\rangle$ as a MPS of $N$ $d$-dimensional physical systems
$|s\rangle$ (e.g. $d=2$ for individual $S=1/2$ spins). We then 
enlarge the Hilbert space by replacing each physical site $|s\rangle$  by two auxiliary
systems $a_k$ and $b_k$ each of a certain dimension $D$ (to be obtained variationally 
to minimize the energy), except for the extremes of the
chain, where only one auxiliary system is added, and set them in the highest
entangled state $|\phi\rangle=\sum_{n=1}^D |n,n\rangle$ (see Fig.
\ref{figverst}). Each of these states is then projected to the real spin by
applying an operator $Q_k$ to each pair to obtain the real entangled system,

\begin{eqnarray}
\nonumber
 |\Psi\rangle &=& \hspace{1cm}Q_1\otimes Q_2\otimes\ldots Q_N
 \hspace{.2cm}|\phi\rangle \ldots |\phi\rangle
 \\
 \label{MPS}
 &=& \sum_{s_1,\ldots,s_N=1}^d {\rm F}_1
 (A^{s_1}_1, \ldots, A^{s_N}_N) |s_1,\ldots,s_N\rangle.
\end{eqnarray}
Here the matrices $A^s_k$ have elements $(A^s_k)_{l,r}=\langle
s|Q_k|l,r\rangle$ and the indices $l$ and $r$ of each
matrix $A^s_k$ are related to the left and right bonds of the
auxiliary systems with their neighbors.
The function F$_1$ denotes the trace of the product of the matrices.
An extension of this state was proposed for two dimensions (see Fig.  \ref{figverst}c), where
now each site is represented as the projection of 4 $D$-dimensional auxiliary systems.
The ground state of a Hamiltonian corresponds to the PEPS with a given dimension $D$ which
minimizes the energy and this is performed by iteratively optimizing each $A$ tensor while
keeping the others fixed, in a similar manner as the finite-size DMRG zipping procedure.
Within this MPS picture, it was shown also by Verstraete and collaborators \cite{verstuli}
that a uniform picture can be given of Wilson's numerical renormalization group method
\cite{wilson} together with White's DMRG.

\section{Time-dependent analysis}

One of the most important recent developments of the DMRG concerns real-time calculations: the so-called
time-dependent DMRG. Its potential applications include the analysis of the evolution of
wave functions, the calculation of
transport properties in low-dimensional systems, non-equilibrium transport through
nanosystems or dissipative quantum mechanics.
A recent comprehensive review on the time-dependent DMRG can be found in Ref. \cite{ulirevt}.

There are two main different schemes to tackle time-dependent problems.
Both different approaches have been classified as {\it dynamic} (Hilbert space) and {\it
adaptive} time-dependent methods. In the first case the Hilbert space for the initial wave
function
is kept as large as possible so as to be able to describe the time-evolved state for
sufficiently long times. In the second approach, instead, the
states kept are calculated at each time interval, adapting to the evolving Hamiltonian.

\begin{figure}[tbp]
\begin{center}
\epsfxsize=4.0in \epsfysize=3.0in \epsffile{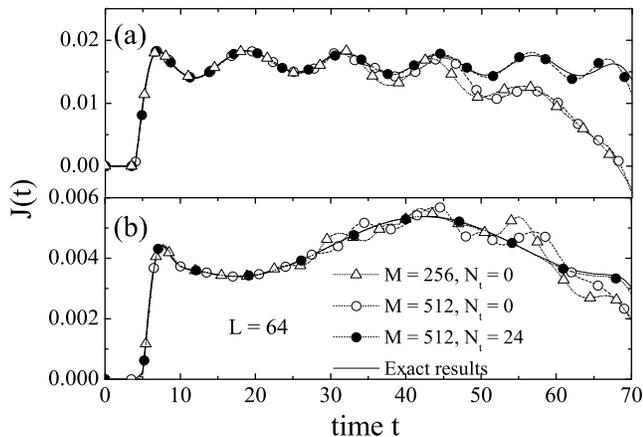}
\end{center}
\caption{Current through a) a non-interacting quantum dot and b) a junction between two
leads (defined in \cite{cazalilla}). $N_t$ is the number of excited target
states, $L$ the number of sites and $M$ the number of states kept (reprinted with 
permission from \cite{luo}). }
\label{figluo}
\end{figure}

The first method was developed by Cazalilla and Marston \cite{cazalilla} in a simplified
version which targeted only the ground state (defined as {\it static} method in
\cite{ulirevt}). After growing
the system up to a certain size, a quantum dot or a junction is added and a time-dependent
perturbation is $H'(t)$ set on. The time-dependent Schr\"odinger equation is then
integrated forward in time using a discretized algorithm:

\begin{equation}\label{eq1}
i \hbar \frac{\partial}{\partial t}| \Psi(t) \rangle = \big[
(H_{\rm trunc} - E_0) + H'(t) \big] |\Psi(t)\rangle\ .
\end{equation}
where the initial state is chosen to be the ground state of the unperturbed system.
As this method does not conserve unitarity, the alternative Crank-Nicholson procedure
\cite{crank},
\begin{equation}
|\Psi(t+dt)\rangle \simeq \frac{1-iH_{\rm trunc}(t)\delta t/2}{1+iH_{\rm
trunc}(t)\delta t/2}|\Psi(t)\rangle
\end{equation}
can be used. The calculation of the denominator can
be calculated using the biconjugate gradient approach\cite{daley}.
If more than one target state is used during the growing procedure, a higher precision is obtained and the
time evolution can be followed longer\cite{luo}.
Fig. \ref{figluo} shows the current through two quantum systems calculated using this
method.

A slightly different approach was used by Schmitteckert\cite{schmitteckert}. Instead of
using the differential equation, he applied the time evolution operator $exp(-iHt)$
directly by using a Krylov basis expansion of the matrix exponential, using a Lanczos
procedure, within the finite-system algorithm. 
Using this method, and keeping information of the evolved wave function in a similar manner
as described below,
transport properties of spinless electrons through strongly interacting 1D systems with arbitrary bias was calculated
\cite{schmitteckert2}. This method does not involve Trotter approximations (see below) and is appropriate
to deal with general geometries and interactions. 
An improvement to this {\it dynamic} method was performed in \cite{feiguin2} by using the 4th order
Runge-Kutta method and targeting a small interval of time (time-step targeted method).

The {\it adaptive} approach\cite{feiguin,daley} was based on the
algorithm developed by
Vidal for the time evolution of weakly entangled quantum states \cite{vidal}, originally
formulated in the matrix product language.

For a one dimensional system with nearest-neighbour interactions, the infinitesimal time evolution operator
$exp(-iHdt)$ can be decomposed using a second order (of course higher orders can also bee applied) 
Trotter-Suzuki expansion into link operators (to order $dt^3$) as:
\begin{equation}
e^{-idt H} \approx e^{-idt H_1/2} e^{-idt H_2/2}\ldots e^{-idt H_2/2}
e^{-idt H_1/2},
\label{}
\end{equation}
where $H_i$ is the Hamiltonian term acting on link $i$.
When applied $t/dt$ times to get one time interval, the error is of the order $(dt^2)t$, so, in the
worst case it scales linearly with $t$ (an extensive error analysis is performed in \cite{gobert2}).

\begin{figure}[ht]
\vspace{9.0cm}
\begin{center}
\epsfxsize=2.0in \epsfysize=0.5in \epsffile{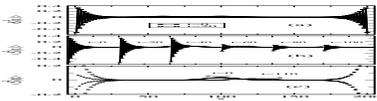}
\end{center}
\caption{Time evolution of a spin excitation wave-packet in a spin-1 Heisenberg
chain with 200 sites. In b) only the 50 leftmost sites are shown. Reprinted from
\cite{feiguin} with permission.}
\label{figfeiguin}
\end{figure}

One important characteristic of the DMRG is that, when operators are calculated acting on
the single sites outside the renormalized blocks, the result is exact. Taking advantage of
this, the above decomposition can be applied sequentially on these exact sites, in a
finite-sweep-like procedure. This algorithm can be easily incorporated into a static DMRG
code.
Using this method, White and Feiguin\cite{feiguin} calculated the same transport properties as in Fig.
\ref{figluo} and their results match the exact result (full line) very precisely for larger times.
They have also calculated other time-dependent responses like the dispersion of a
spin excitation in a 200-sites spin-1 Heisenberg chain (see Fig. \ref{figfeiguin}).

The results obtained by this method are very accurate thus allowing to obtain
frequency-dependent quantities by Fourier transformation\cite{feiguin}. Some
recent applications include the calculation of the evolution of one-dimensional
density waves\cite{kollath} and the dynamics of the superfluid-to-Mott transition\cite{clark} 
(here using the matrix formalism\cite{vidal}) in ultracold bosons in an optical lattice
 and the analysis of charge spin separation in cold Fermi gases\cite{ulisep} by
observing the time evolution of wave packets\cite{jagla}. It was also applied
to calculate the Zener breakdown in band and Mott insulators\cite{oka}, 
real-time dynamics of spin 1/2 Heisenberg chains\cite{gobert2}, conductance through strongly correlated
systems\cite{alhassanieh} and a single atom transistor in 1D optical lattice\cite{micheli}.
An interesting application of this time-dependent analysis is the recent study of
the evolution of the entropy of a block of spins in the Heisenberg chain after a sudden quench in
the anisotropy\cite{dechiara}, which gives information of the propagation of entanglement and critical behaviour.

This method is very precise and easy to implement. However, it has the disadvantage of
being restricted by geometry and nearest-neighbour interactions. For other geometries the 
{\it dynamic} method mentioned above is preferred.

\section{Zero temperature dynamics}

DMRG can also be used to calculate dynamical properties of low-dimensional
systems, useful to interpret experimental results from, for example, nuclear
magnetic resonance (NMR), neutron scattering, optical absorption and 
photoemission, among others. There have been two main approaches to the
dynamics: the Lanczos \cite{karendin,kuehner} and correction vector
techniques \cite{kuehner,ramasesha,jeckelmann}. The first gives complete
information of the whole excitation spectrum at the expense of less accuracy
for large systems, specially at high energies. The latter, instead, focuses
on particular energy values and gives more precise information, being
numerically much more expensive.

\subsection{Lanczos technique}
\label{sec:lanczos}

We want to calculate the following dynamical correlation function at $T=0$:
\begin{equation}
C_{A}(t-t^{\prime })=\langle \psi _{0}|A^{\dagger }(t)A(t^{\prime })|\psi
_{0}\rangle ,  \label{eq:ca}
\end{equation}
where $A^{\dagger }$ is the Hermitean conjugate of the operator $A$, $A(t)$
is the Heisenberg representation of $A$, and $|\psi _{0}\rangle $ is the
ground state of the system. Its Fourier transform is:
\begin{equation}
C_{A}(\omega )=\sum_{n}|\langle \psi _{n}|A|\psi _{0}\rangle |^{2}\;\delta
(\omega -(E_{n}-E_{0})),
\end{equation}
where the summation is taken over all the eigenstates $|\psi _{n}\rangle $
of the Hamiltonian $H$ with energy $E_{n}$, and $E_{0}$ is the ground state
energy.

Defining the Green's function
\begin{equation}  \label{eq:din}
G_A(z)=\langle \psi_0 | A^{\dagger}(z-H)^{-1} A |\psi_0 \rangle,
\end{equation}
the correlation function $C_A(\omega)$ can be obtained as
\begin{equation}
C_A(\omega)=-\frac{1}{\pi}\lim_{\eta\to 0^+}\mathrm{Im} \; G_A(\omega+i\eta
+E_0).
\end{equation}

The function $G_A$ can be written in the form of a continued fraction:
\begin{equation}  \label{eq:frac}
G_A(z)=\frac{\langle \psi_0 | A^{\dagger} A|\psi_0\rangle}{z-a_0-\frac{b_1^2%
} {z-a_1-\frac{b_2^2}{z-...}}}
\end{equation}
The coefficients $a_n$ and $b_n$ can be obtained using the following
recursion equations \cite{carlos,proj}:
\begin{equation}
|f_{n+1}\rangle =H|f_n\rangle -a_n|f_n\rangle -b_n^2|f_{n-1}\rangle
\end{equation}
where
\begin{eqnarray}
|f_0\rangle &=& A|\psi_0\rangle  \nonumber \\
a_n&=&\langle f_n|H|f_n\rangle/\langle f_n|f_n\rangle,  \nonumber \\
b_n^2&=&\langle f_n|f_n\rangle/\langle f_{n-1}|f_{n-1}\rangle; \;\; b_0=0
\end{eqnarray}

For finite systems the Green's function $G_A(z)$ has a finite number of
poles. Within this formulation it is not necassary to compute all coefficients
$a_n$ and $b_n$ since the spectrum is nearly converged already with the first few
tens or hundreds of them, depending on the problem.
The DMRG technique presents a good framework to calculate such
quantities. With it, the ground state, Hamiltonian and the operator $A$
required for the evaluation of $C_A(\omega)$ are obtained. An important
requirement is that the reduced Hilbert space should also describe with
great precision the relevant excited states $|\psi_n \rangle $. This is
achieved by choosing the appropriate target states. For most systems it is
enough to consider as target states the ground state $|\psi_0\rangle$ and
the first few $|f_n\rangle $ with $n=0,1...$ and $|f_0\rangle=
A|\psi_0\rangle$ as described above. In doing so, states in the reduced
Hilbert space relevant to the excited states connected to the ground state
via the operator of interest $A$ are included. The fact that $|f_0\rangle$
is an excellent trial state, in particular, for the lowest triplet
excitations of the two-dimensional antiferromagnet was shown in Ref.~\cite
{linden}. Of course, if the number $m$ of states kept per block is fixed,
the more target states considered, the less precisely each one of them is
described. An optimal number of target states and $m$ have to be found for
each case. Due to this reduction, the algorithm can be applied up to certain
lengths, depending on the states involved. For longer chains, the higher
energy excitations will become inaccurate. Proper sum rules have to be
calculated to determine the errors in each case.

\begin{figure}[tbp]
\vspace{2cm}
\begin{center}
\epsfxsize=3.0in \epsfysize=2.0in 
\epsffile{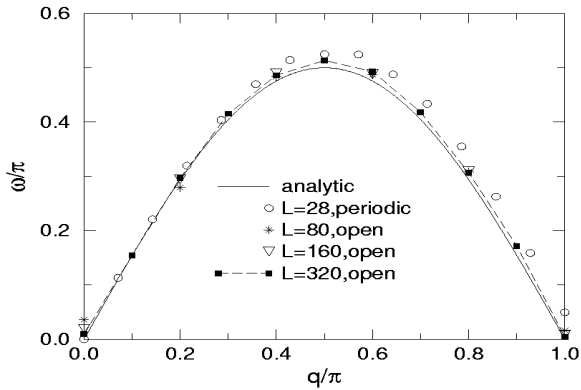}
\end{center}
\caption{Lower bound of the continuous excitation band of the AF $S=1/2$ Heisenberg model
for different system sizes $L$
(reprinted from Ref. \protect\cite{kuehner} with permission). The full line is the
analytical result for the infinite system using Bethe Ansatz.}
\label{figdin}
\end{figure}

This method has been successfully applied to a number of problems, like spin
1/2 (see Fig.\ref{figdin}), 3/2 \cite{karen12,karen32,karendin,kuehner} and spin 1
chains\cite{kuehner}, the
spin-boson model \cite{nishiyama}, the Holstein model \cite{zhangjeck,matsuedaholstein} and
spin-orbital chains in external fields \cite{yuhaas}. It was also applied to
extract spin-chain dispersion relations \cite{okunishi}, dynamics of spin
ladders \cite{nunner}, spectral functions in the infinite-U Hubbard model
\cite{penc}, optical response and spinon and holon excitations in 1D Mott insulators
\cite{kancharla,matsueda}. In a recent work, this method, together with the correction vector
described below, was used to calculate the spectral function of ultrasmall Kondo systems
and analyzed finite-size and even-odd effects\cite{monien2} with great accuracy. 
In Section 10 we will describe its application as the impurity solver within the
Dynamical Mean Field Theory method (DMFT).

\subsection{Correction-vector method}

This method focuses on a particular energy or energy window, allowing for a
more precise description in that range and the possibility of calculating
spectra for higher energies. Instead of using the tridiagonalization of the
Hamiltonian, but in a similar spirit regarding the important target states
to be kept, the spectrum can be calculated for a given $z=w+i\eta$ by using
a correction vector (related to the operator $A$ that can depend on momentum
$q$).

Following the Green's function given above, the (complex) correction vector $%
|x(z)\rangle$ can be defined as:
\begin{equation}
|x(z)\rangle = \frac{1}{z-H}A |\psi_0 \rangle
\end{equation}
so the Green's function can be calculated as $G(z)=\langle \psi_0
|A^{\dagger} |x(z)\rangle$. 

Separating the correction vector in real and imaginary parts $|x(z)\rangle =
|x^r(z)\rangle + i |x^i(z)\rangle$ we obtain
\begin{eqnarray}
((H-w)^2 + \eta^2)|x^i(z)\rangle &=& -\eta A |\psi_0 \rangle  \nonumber \\
|x^r(z)\rangle &=& \frac{1}{\eta}(w-H)|x^i(z)\rangle
\label{eq:cv}
\end{eqnarray}
The former equation is solved using the conjugate gradient method. In order
to keep the information of the excitations at this particular energy the
following states are targeted in the DMRG iterations: The ground state $%
|\psi_0 \rangle$, the first Lanczos vector $A |\psi_0 \rangle$ and the
correction vector $|x(z)\rangle$. Even though only a certain energy is
focused on, DMRG gives the correct excitations for an energy range
surrounding this particular point so that by running several times for
nearby frequencies, an approximate spectrum can be obtained for a wider
region \cite{kuehner}.

\begin{figure}[tbp]
\vspace{4cm}
\begin{center}
\epsfxsize=2.5in \epsfysize=1.7in \epsffile{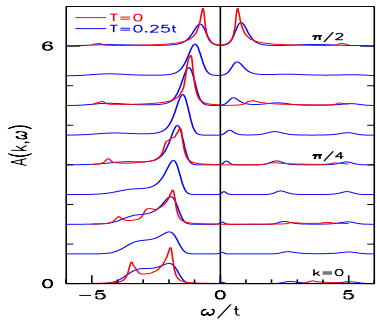}
\end{center}
\caption{Single-particle spectral function for the 1D Hubbard insulator for $U=4t$ and several momenta.
The black curves are the DMRG results for $T=0$ (64 sites) and
the blue curve corresponds to Quantum Monte Carlo data for $T=0.25t$
(reprinted from Ref. \protect\cite{matsueda} with permission).}
\label{fig:matsueda}
\end{figure}

A variational formulation of the correction vector technique has been
developed in \cite{jeckelmann}. In order to solve Eq. \ref{eq:cv}, the following
equation is minimized w.r.t $|X\rangle$:
\begin{equation}
W_{A,\eta}(\omega,X)=\langle X|(H-w)^2 + \eta^2|X\rangle + \eta \langle \psi_0|A |X \rangle
+ \eta \langle X|A |\psi_0 \rangle .
\label{eq:jeck}
\end{equation}
For any $\eta \ne 0$ and finite $\omega$ this function has a well defined minimum for the
quantum state which is solution of (\ref{eq:cv}), {\it i.e.} $ |x^i(z)\rangle$. This method 
is
very similar to thge correction vector one, but has a smaller error in the determination of the spectral function
\cite{jeckreview}.
It has been successfully applied to
calculate the optical conductivity of Mott insulators \cite
{essler,jeck03,kim}, 
spectral functions of the 1D
Hubbard insulator \cite{matsueda} (see Fig. \ref{fig:matsueda}) and away from half-filling\cite{benthien}  
and, more recently, quantum impurity spectral functions for one\cite{nishimotojeck} and two impurities\cite{nishimoto2imp}.

The correction vector model has also been applied to determine the nonlinear
optical coefficients of Hubbard chains and related models \cite{pati}, to
calculate ac conductivity of the Bose-Hubbard model \cite{kuehner2} and the
single-impurity Anderson model.\cite{raas} It has also been recently applied to 
calculate the conductance through an interacting spinless fermionic systems in the linear
response using the Kubo formula\cite{wolfle}. 

An interesting development for calculating response functions in single impurity
systems in the presence of a magnetic field was done in \cite{hofstetter} by
using the DMRG within Wilson's NRG to obtain the Green's function.
Very recently, Verstraete and collaborators proposed a variational method (by optimizing 
the correction vector) for evaluating
Green's functions applicable to 1D quantum chain models, based on the matrix product states
formalism and successfully applied to the single impurity Anderson model \cite{verstuli}.

\section{Classical systems}

The DMRG has been very successfully extended to study classical systems. For
a detailed description we refer the reader to Ref. \cite{nishino}. Since 1D
quantum systems are related to 2D classical systems\cite{clasico}, it is
natural to adapt DMRG to the classical 2D case. This method is based on the
renormalization group transformation for the transfer matrix $T$ (TMRG)\cite
{bursill96}. It is a variational method that maximizes the partition
function using a limited number of degrees of freedom, where the variational
state is written as a product of local matrices\cite{ostlund}. For 2D
classical systems, this algorithm is superior to the classical Monte Carlo
method in accuracy, speed and in the possibility of treating much larger
systems. A recent improvement of this method considering periodic boundary
conditions is given in \cite{gendiar2} and a detailed comparison between
symmetric and asymmetric targeting is done in \cite{shibata2}. TMRG has
also been successfully used to renormalize stochastic transfer matrices in a
study of cellular automatons\cite{stochastic}. The calculation of
thermodynamical properties of 3D classical statistical systems has been
proposed\cite{nishino5} where the eigenstate of the transfer matrix with
maximum eigenvalue is represented by the product of local tensors optimized
using DMRG.

A further improvement to this method is based on the corner transfer matrix
\cite{baxter}, the CTMRG\cite{okunishi3,takasaki,nishino3,ritter,ueda,ueda2} and can be
generalized to any dimension\cite{okunishi2}. In Ref. \cite{okunishi4}, there is
an interesting analysis on the low effective theory underlying the DMRG, based on
a real-space RG of the corner Hamiltonian and applied to spin chains.

It was first applied to the Ising \cite{nishino,drz,drz2,kaulke} and
Potts models\cite{carlon}, where very accurate density profiles
and critical indices were calculated. Further applications have included
non-hermitian problems in equilibrium and non-equilibrium physics. In the
first case, transfer matrices may be non-hermitian and several situations
have been considered: a model for the Quantum Hall effect\cite{kondev}, the $%
q$-symmetric Heisenberg chain related to the conformal series of critical
models\cite{peschel} and the anisotropic triangular nearest and next-nearest
neighbour Ising models\cite{gendiar3}. In the second case, the adaptation of
the DMRG to non-equilibrium physics like the asymmetric exclusion problem
\cite{hieida2} and reaction-diffusion problems \cite{peschel1,carlon1} has
shown to be very successful. It has also been applied to stochastic lattice
models like in \cite{jef} and to the 2D XY model \cite{chung3}, to the study
of simplified models for polimerized membranes in thermal equilibrium (folding of triangular
lattices) \cite{nishiyama4}.

\section{Finite-temperature DMRG}

The first attempt to include the effect of finite temperature in the DMRG
procedure was performed by Moukouri and Caron\cite{moukouri}. They considered the
standard
DMRG taking into account several low-lying target states (see Eq.~\ref{eq:pl}%
) to construct the density matrix, weighted with the Boltzmann factor ($\beta
$ is the inverse temperature):
\begin{equation}  \label{eq:pl2}
\rho_{ii^{\prime}}=\sum_l e^{-\beta E_l} \sum_j \phi_{l,ij}
\phi_{l,i^{\prime}j}
\end{equation}
With this method they performed reliable calculations of the magnetic
susceptibility of quantum spin chains with $S=1/2$ and $3/2$, showing
excellent agreement with Bethe Ansatz exact results. They also calculated
low temperature thermodynamical properties of the 1D Kondo Lattice Model\cite
{moukouri3} and of organic conductors \cite{moukouri4}. Zhang et al.\cite
{zhang} applied the same method in the study of a magnetic impurity embedded
in a quantum spin chain.

An alternative way of incorporating temperature, called 
the Transfer Matrix DMRG (TDMRG) stems from the Trotter-Suzuki
expansion of the partition function of a one-dimensional model, turning it
into a classical two-dimensional model, with the new axis corresponding to imaginary time (inverse temperature).
The extension of the DMRG method to classical 
systems paved the way to
study 1D quantum systems at non zero temperature\cite{bursill2,trotter,xiangwang,ueda2,shibata}(see
also a recent review in Ref.\cite{shibatareview}). In this case the system
is infinite and the finiteness is in the level of the Trotter approximation.
The free energy in the thermodynamic limit is determined by the largest eigenvalue of the
transfer matrix. Being a high-temperature
expansion, the method loses precision at low temperatures.

Very nice results have been obtained for the dimerized, $S=1/2$, XY model,
where the specific heat was calculated involving an extremely small basis set
\cite{bursill2} ($m=16$), the agreement with the exact solution being much
better in the case where the system has a substantial gap. It has also been
used to calculate thermodynamic properties of the a\-ni\-so\-tropic $S=1/2$
Heisenberg model, with relative errors for the spin susceptibility of less
than $10^{-3}$ down to temperatures of the order of $0.01J$ keeping $m=80$
states\cite{xiangwang}. A complete study of thermodynamic properties like
magnetization, susceptibility, specific heat and temperature dependent
correlation functions for the $S=1/2$ and 3/2 Heisenberg models was done in
\cite{xiangt}. Other applications have been the calculation of the
temperature dependence of the charge and spin gap in the Kondo insulator\cite
{ammon1}, the calculation of thermodynamic properties of ferrimagnetic
chains\cite{maisinger} and spin ladders\cite{wanglu}, the study of impurity
properties in spin chains\cite{rommer,maruyama1}, frustrated quantum spin
chains\cite{maisinger2}, t-J\cite{ammon} and spin ladders\cite{naefwang} and
dimerized frustrated Heisenberg chains\cite{klumper}. Recent studies include the thermodynamics
and crossover phenomena in the 1D t-J model using this method which led to very accurate results, as
compared to the particular case of the supersymmetric limit where analytic Bethe-Ansatz
results are available\cite{sirker} and the calculation of the specific heat in doped 
anisotropic Hubbard ladders with charge-order instability\cite{ohta}. 

Recently, new proposals for calculating finite-temperature properties arose based on 
the imaginary-time evolution of the wave function in the matrix-product\cite{verst2,zwolak,verstripoll} 
and ancillary states \cite{verst2,feiguintemp} formalisms to simulate
evolutions of physical quantities in real and imaginary time and at finite
temperature. They are  based on the so-defined {\it projected entangled-pair
states} (PEPS) and the matrix product density operators (MPDO). The evolution in
imaginary time to construct thermal and mixed states leads to a versatile DMRG algorithm that 
is restricted neither to large temperatures nor to homogeneous systems and opens the possiblity of
simulating finite temperature, dissipation and decoherence effects.
From the DMRG point of view, the PEPS or ancilla method is particularly appropriate. 
The Hilbert space is enlarged by introducing auxiliary states
(in the form of an extra chain coupled to the system for example) and then the imaginary-time evolution 
of a pure state in the enlarged space is calculated (this process is known as ``purification" in quantum information theory
\cite{nielsenchuang}).
The ancilla sites serve as a thermodynamic bath
which is traced out to obtain the thermodynamic behaviour of the original system.

\begin{figure}[htbp]
\begin{center}
\epsfxsize=2.5in \epsfysize=1.7in \epsffile{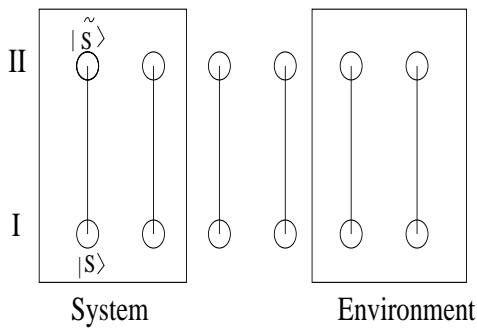}
\end{center}
\caption{Schematic representation of the ancilla approach used to calculate imaginary-time 
evolution. The real system corresponds to the lower sites (I) and the ancillary 
states are labelled by II. Full lines represent maximally entangled states.} 
\label{figancilla}
\end{figure}

\begin{figure}[tbp]
\begin{center}
\epsfxsize=3.0in \epsfysize=2.0in \epsffile{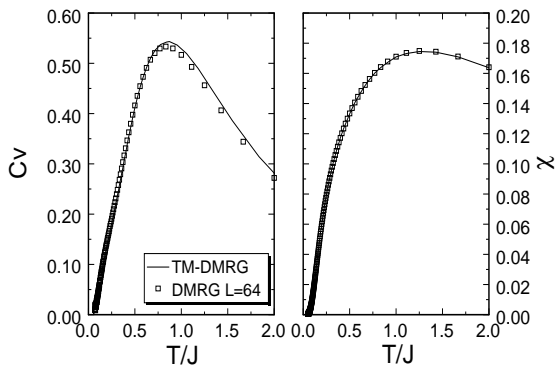}
\end{center}
\caption{Specific heat and magnetic susceptibility of the Haldane chain ($S=1$) with 64 sites  
calculated using the Trotter-Suzuki time-evolution algorithm. Also shown are the results using 
Transfer Matrix DMRG (reprinted from Ref. \protect\cite{feiguintemp} with permission).}
\label{fig:feiguintemp}
\end{figure}

Let our original system be defined by states $|s_i\rangle$ 
 and our auxiliary bath by $|\tilde{s_i}\rangle$ (Fig. \ref{figancilla}).
Following the notation of \cite{feiguintemp}, we define the non-normalized {\it pure} state
\begin{equation}
|\psi(\beta)\rangle= e^{-\beta H/2}|\psi(0)\rangle
\end{equation}
where the Hamiltonian $H$ acts only on the real system (I).
If $\beta=0$ ($T\to \infty$) this wave function is in a maximally entangled state between
the original and auxiliary states:
\begin{equation}
|\psi(0)\rangle= \prod_{i}\sum_{s_i} |s_i \tilde{s_i}\rangle
\end{equation}
The partition function for arbitrary $\beta$ is $S=\langle \psi(\beta)|\psi(\beta)\rangle$ and the thermal average of an operator
$A$ is calculated as
\begin{equation}
\langle A(\beta)\rangle= \frac {\langle \psi|A|\psi\rangle }{\langle \psi|\psi\rangle}.
\end{equation}
To obtain the temperature-dependent wave function, one can start with the maximally entangled state
$|\psi(0)\rangle$ and then evolve in imaginary time applying the Hamiltonian to the real system only. 
Within the DMRG framework, the blocks can be thought of as for example shown in Fig.\ref{figancilla},
where each pair $|s_i \tilde{s_i}\rangle$ forms a supersite and the maximally entangled state is represented 
as a full line. The imaginary-time evolution is performed with any of the real-time algorithms
(for example, the {\it adaptive} algorithm \cite{feiguin,daley} for nearest neighbour Hamiltonians).
In \cite{feiguintemp}, the specific heat, magnetic susceptibility and the correlation length are calculated
in spin chains with an accuracy comparable to the transfer-matrix DMRG 
(see Fig.\ref{fig:feiguintemp}).

\subsection{Finite-temperature dynamics}

In order to include temperature in the calculation of dynamical quantities,
the Transfer Matrix RG described above (TMRG) \cite{bursill2,xiangwang,shibata}
was extended to obtain imaginary time correlation functions\cite
{wangbook,mutou,naef}. After Fourier transformation in the imaginary time
axis, analytic continuation from imaginary to real frequencies is done using
maximum entropy (ME). The combination of the TMRG and ME is free from
statistical errors and the negative sign problem of Monte Carlo methods but still
has the extrapolation error of the analytic continuation.
Since we are dealing with the transfer matrix, the thermodynamic limit (infinite system size) 
can be discussed directly without extrapolations. However, in the present
scheme, only local quantities can be calculated.

A systematic investigation of local spectral functions is done in Ref. \cite
{naef} for the anisotropic Heisenberg antiferromagnetic chain. The authors
obtain good qualitative results especially for high temperatures but a
quantitative description of peaks and gaps are beyond the method, due to the
severe intrinsic limitation of the analytic continuation. This method was
also applied with great success to the 1D Kondo insulator\cite{mutou} where the
temperature dependence of the local density of states and local dynamic spin
and charge correlation functions were calculated. A modfication of this method
to avoid the use of the ill-posed analytic continuation was done in \cite{sirker2}
by considering a path-integral approach to calculate real-time correlation functions at finite
temperatures.

\section{Dynamical Mean Field Theory using DMRG}

The Dynamical Mean Field Theory (DMFT) has become one of the basic methods
to calculate realistic electronic band structure in strongly correlated
systems \cite{pt}. At the heart of the DMFT method is the solution of
an associated quantum impurity model where the environment of the impurity
has to be determined self-consistently. Therefore the ability to obtain
reliable DMFT solutions of lattice model Hamiltonians relies directly on the
ability to solve quantum impurity models. Among the \textit{a priori} exact
numerical algorithms available we find the Hirsch-Fye Quantum Monte Carlo
\cite{hf,mr} method and Wilson's Numerical Renormalization Group (NRG)\cite
{wilson,bulla,vollhardt}. While the former, a finite-temperature method, is
very stable and accurate at the Matsubara frequencies, its main drawback is
the access to real frequency quantities for the calculation of spectral
functions which requires less controlled techniques for the analytic
continuation of the Green functions. The second method can be formulated
both at $T$=0 and finite (small) $T$ and provides extremely accurate results
at very small frequencies, at the expense of a less accurate description of
the high energy features.

In order to overcome the difficulties encountered by these other methods, we
will show that DMRG can be used very reliably to solve the related impurity
problem within DMFT\cite{garciadmft}. By using the DMRG to solve the 
impurity, no \textit{a priori} approximations are made and the
method provides equally reliable solutions for both gapless and gapful
phases. More significantly, it provides accurate estimates for the
distributions of spectral intensities of high frequency features such as the
Hubbard bands, that are of main relevance for analysis of X-ray
photoemission and optical conductivity experiments.

We will now very briefly describe the method applied to the Mott transition in
the Hubbard model. The Hamiltonian of the Hubbard model is defined by Eq.
\ref{Hub}.
Applying DMFT to this model leads to a mapping of the original lattice model
onto an associated quantum impurity problem in a self-consistent bath. In
the particular case of the Hubbard model, the associated impurity problem is
the single impurity Anderson model (SIAM), where the hybridization function $%
\Delta(\omega)$, which in the usual SIAM is a flat density of states of the
conduction electrons, is now to be determined self-consistently. More
precisely, for the Hamiltonian (\ref{Hub}) defined on a Bethe lattice of
coordination $d$, one takes the limit of large $d$ and exactly maps the
model onto a SIAM impurity problem with the requirement that $\Delta(\omega)
= t^2 G(\omega)$, where $G(\omega)$ is the impurity Green's function. At the
self-consistent point $G(\omega)$ coincides with the \emph{local} Green's
function of the original lattice model \cite{review}. We take the
half-bandwidth of the non interacting model as unit of energy, $t=1/2$.

The Green's function of the impurity problem is an important quantity in
this algorithm: \cal{ G}$_0(\omega) = 1/ (\omega +\mu -
\Delta(\omega)) = 1/ (\omega +\mu - t^2 G(\omega))$. Thus, to implement the
new algorithm we shall consider \cite{qimiao,mpl} a general representation
of the hybridization function in terms of continued fractions that define a
\emph{parametrization} of $\Delta(\omega)$ in terms of a set of real and
positive coefficients. Since it is essentially a Green's function, $\Delta(z)
$ can be decomposed into ``particle'' and ``hole'' contributions as $\Delta
(z)=\Delta^>(z)+\Delta^<(z)$ with $\Delta^>(z)=t^2\langle gs|c \frac{1}{%
z-(H-E_0)} c^{\dagger} |gs \rangle$ and $\Delta^<(z)=t^2\langle
gs|c^{\dagger}\frac{1}{z+(H-E_0)} c |gs \rangle$ for a given Hamiltonian, H
with ground-state energy $E_0$. By standard Lanczos technique, H can be 
tridiagonalized and the functions $\Delta^>(z)$ and $\Delta^<(z)$
can be expressed in terms of respective continued fractions\cite{karendin}.
As first implemented in Ref.\cite{qimiao,mpl}, each continued fraction can
be represented by a chain of auxiliary atomic sites whose energies and
hopping amplitudes are given by the continued fraction diagonal and
off-diagonal coefficients respectively.

As a result of the self-consistency condition, the two chains representing
the hybridization, are ``attached'' to the right and left of an atomic site
to obtain a new SIAM Hamiltonian, H. In fact \cal{G}$_0(z)$ constitutes
the local Green's function of the site plus chain system. The algorithm in
Ref.\cite{qimiao,mpl}, basically consists in switching on the local Coulomb
interaction at the impurity site of the SIAM Hamiltonian and using the Lanczos
technique to re-obtain $\Delta(z)$, iterating the procedure until the set of
continued fractions coefficients converges. By using the DMRG, the number of
auxiliary sites that can be used in the hybridization chains is much larger
than in the exact diagonalization scheme, leading to more accurate results
\cite{gebhard}. An alternative way of using DMRG to solve the impurity
problem, which does not rely on the continued fraction expansion was
developed in \cite{jeckelmandmft}.

The SIAM Hamiltonian therefore reads
\begin{eqnarray}
H &=& \sum_{{\sigma,\alpha=-N_C; \alpha\ne0}}^{N_C} a_{\alpha}
c_{\alpha\sigma}^{\dagger} c_{\alpha\sigma} +\sum_{{\sigma,\alpha=-(N_C-1);
\alpha\ne0,-1}}^{N_{C}-1} b_\alpha (c_{\alpha\sigma}^{\dagger} c_{
\alpha+1\sigma}+h.c. )  \nonumber \\
&&+\sum_{\sigma,\alpha=\pm1} b_0 (c_{\sigma}^{\dagger} c_{\alpha\sigma} +
h.c.) + U (n_{ \uparrow}-\frac{1}{2}) (n_{ \downarrow}-\frac{1}{2})
\end{eqnarray}
with $c_{\sigma}$ being the destruction operator at the impurity site, and $%
c_{\alpha\sigma}$ being the destruction operator at the $\alpha$ site of the
hybridization chain of $2 N_C$ sites. The set of parameters $\{ a_{\alpha},
b_\alpha \}$ are directly obtained from the coefficients of the continued
fraction representations of $\Delta(z)$ by the procedure just described.

\begin{figure}[ht]
\begin{center}
\epsfxsize=3.0in \epsfysize=2.5in
\epsffile{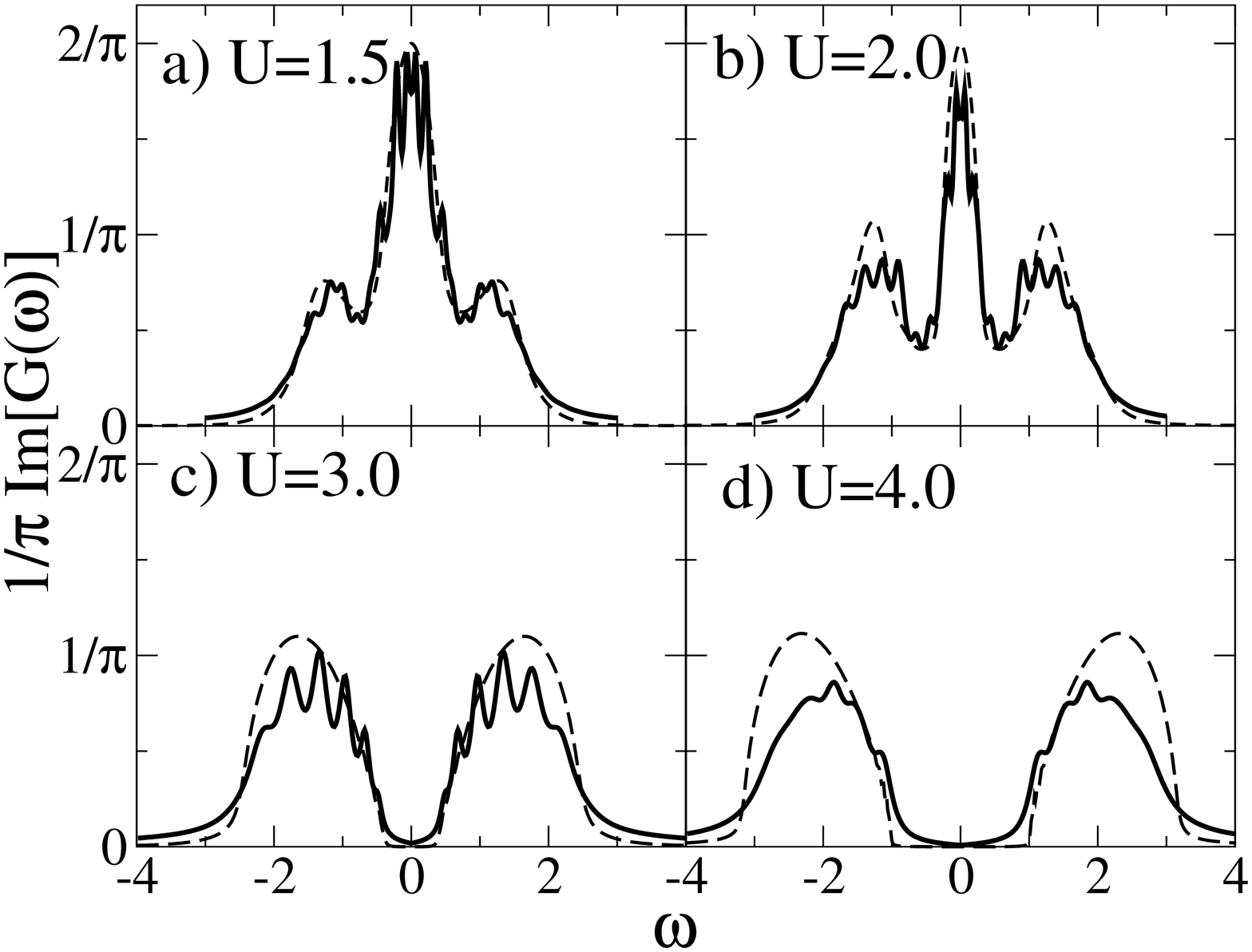}
\caption{Density of states
of the half-filled Hubbard Model. We also show the IPT results (dashed
lines) (see text). }
\label{figdmft}
\end{center}
\end{figure}

In Fig.\ref{figdmft} we show the DMFT+DMRG results (solid lines) for the
density of states (DOS) for several values of increasing interaction U. The
results are compared to the Iterated Perturbation Theory (IPT) results
(dashed lines)\cite{gk,zrg}.

Results for the imaginary part of the Green's funtions on the Matsubara
frequencies match the precise Monte Carlo solutions at low temperatures. We
also obtain accurate values for the two distinct critical values of the
interaction $U_{c1}==2.39 \pm 0.02$ and $U_{c2}=3.0 \pm 0.2$. To this end,
reaching larger system sizes turned out to be important in order to perform
proper extrapolations and overcome finite-size effects.

As a conclusion, using DMRG as the impurity-solver of the DMFT, increases its performance.
Large systems can be
considered and accurate values of the critical interactions are obtained in
agreement with NRG predictions allowing for a non-trivial test of the
accuracy of this method. In contrast with NRG, however, this new algorithm
deals with all energy scales on equal footing which allowed us to find
interesting substructure in the Hubbard bands of the correlated metallic
state. The ability of the new algorithm to directly deal with the high
energy scales is a very important feature which is relevant for the
interpretation of high resolution photoemission spectroscopies.\cite{pt} In
addition, with this method, realistic band-structure calculations of systems
with a larger number of degrees of freedom  and other dopings away from half-filling can be handled 
\ref{garciamiranda}.

\section{Summary and outlook}

The DMRG method has proven to be a very reliable and versatile
numerical method that can be applied to a broad spectrum of problems in
physics in a variety of fields such as Condensed Matter,
Statistical Mechanics, High Energy and Nuclear Physics, Quantum Chemistry and Quantum Information Theory.
It is nowadays recognized as one of the most accurate and efficient numerical techniques.
We have reviewed the most recent developments to DMRG among which are the very accurate
real-time applications, the use of quantum information concepts which give an interesting
perspective and the possibility of using it as the impurity-solver in DMFT calculations.
These new improvements have triggered a great activity and numerous papers have been published
applying these new techniques.

However, the DMRG is still evolving and presents very interesting potential new applications.
Among the fields where DMRG has still to be exploited are the time-dependent analysis of
quantum and dissipative systems, the application in DMFT band-structure calculations of more complex 
and realistic systems and quantum chemistry.
The connection of DMRG with quantum information theory paves the way to an important new
field of application and, very possibly, to dimensions higher than one.

In view of the fact that numerical simulations constitute an essential tool
in physics and quantum chemistry, a group of experts have started an
interesting initiative called the ALPS (Algorithms and Libraries for Physics
Simulations) project \cite{alps}. It consists of an open source of
optimized software aimed at studying strongly correlated systems such as quantum magnets,
strongly correlated fermionic systems and lattice bosons where methods like classical and quantum
Monte Carlo, exact diagonalization and DMRG are included.

\section*{Acknowledgments}

The author is grateful to numerous fruitful discussions on DMRG with most of the people involved with this method
in particular with P. Horsch (who drew her attention to the 
method in its very early stages), S. White, I. Peschel, X. Wang, T. Xiang, T. Nishino, 
M. A. Mart\'{\i}n-Delgado, G. Sierra, S. Ramasesha, E. Jeckelmann, R. Noack, 
U. Schollw\"ock, P. Schmitteckert, D. Garc\'{\i}a and M. Rozenberg.

The author is Staff Researcher of the National Council for Science and Technology,
CONICET (Argentina) and Fellow of the Guggenheim Foundation (2005). This project was performed under grants PICT 02 03-12742 
and PICT 03-13829 of the ANPCyT (Argentina).



\begin{thebibliography}{12}
\bibitem{white1}  S. White, Phys. Rev. Lett. \textbf{69}, 2863 (1992)

\bibitem{book}  \textit{Density Matrix Renormalization}, edited by I.
Peschel, X. Wang, M. Kaulke and K. Hallberg (Series: Lecture Notes in
Physics, Springer, Berlin, 1999)

\bibitem{scholl}  A. Schollw\"{o}ck, \textit{The density matrix
renormalization group}, Rev. Mod. Phys.,{\textbf 77}, 259, Jan. 2005 (see also
R. Noack and S. Manmana, AIP Conf. Proc. {\bf 789}, 93 (2005) for a more general review on
numerical methods for interactin quantum systems).

\bibitem{white2}  S. White, Phys. Rev. B \textbf{48}, 10345 (1993)

\bibitem{braychui}  J. Bray and S. Chui, Phys. Rev. B \textbf{19}, 4876
(1979)

\bibitem{panchen}  C. Pan and X. Chen, Phys. Rev. B \textbf{36}, 8600
(1987); M. Kovarik, Phys. Rev. B \textbf{41}, 6889 (1990)

\bibitem{xiangghering}  T. Xiang and G. Gehring, Phys. Rev. B \textbf{48},
303 (1993)

\bibitem{wilson}  K. Wilson, \textit{Rev. Mod. Phys.} \textbf{47}, 773 (1975)

\bibitem{verstlatorre} F. Verstraete et al, Phys. Rev. Lett. {\bf 94}, 140601 (2005)

\bibitem{noackwhite}  S. White and R. Noack, Phys. Rev. Lett. \textbf{68},
3487, (1992); R. Noack and S. White, Phys. Rev. B \textbf{47}, 9243 (1993)

\bibitem{gaite}  J. Gaite, Mod. Phys. Lett. A\textbf{16}, 1109 (2001)


\bibitem{feynman}  R. Feynman, \textit{Statistical Mechanics: A Set of
Lectures}, (Benjamin, Reading, MA, 1972)

\bibitem{whitebook}  R. Noack and S. White in Ref.~\cite{book}, Chap. 2(I)

\bibitem{sierraparticle}  M. Mart\'{\i}n-Delgado, G. Sierra and R. Noack, J.
Phys. A: Math. Gen. \textbf{32}, 6079 (1999)


\bibitem{scalapino}  S. R. White, I. Affleck and D. Scalapino, Phys. Rev. B
\textbf{65}, 165122 (2002)

\bibitem{lanczos}  See E. Dagotto, Rev. Mod. Phys. \textbf{66}, 763 (1994); N.  Laflorencie and D. Poilblanc, 
Lect. Notes Phys. {\bf 645}, 227 (2004)

\bibitem{davidson}  E. R. Davidson, J. Comput. Phys. \textbf{17}, 87 (1975);
E.R. Davidson, \textit{Computers in Physics} \textbf{7}, No. 5, 519 (1993).

\bibitem{cavo}  S. White, Phys. Rev. Lett. \textbf{77}, 3633 (1996)

\bibitem{white4}  T. Nishino and K. Okunishi, Phys. Soc. Jpn. \textbf{64},
4084 (1995); U. Schollw\"{o}ck, Phys. Rev. B \textbf{58}, 8194 (1998) and
Phys. Rev. B \textbf{59}, 3917 (1999) (erratum).

\bibitem{preskill} J. Preskill, J. Mod. Opt. {\bf 47}, 127 (2000) (quant-ph/9904022)


\bibitem{osborne}  T. J. Osborne and M. A. Nielsen, Quantum Information
Processing, Volume 1, Issue 1-2, 45 (2002) (available on-line:
http://www.kluweronline.com/issn/1570-0755/); ibid. Phys. Rev. A {\bf 66}, 032110 (2002);
F. Verstraete, M. Popp and J. I. Cirac, Phys. Rev. Lett. {\bf 92}, 027901 (2004)

\bibitem{verst2}  F. Verstraete and J. Cirac, preprint, cond-mat/0407066

\bibitem{verstraete}  F.Verstraete, D. Porras and C. Cirac, Phys. Rev. Lett. \textbf{93}
227205 (2004); F. Verstraete, M. A. Mart\'{\i}n-Delgado and J. I. Cirac, Phys. Rev. Lett.
{\bf 92}, 087201 (2004)

\bibitem{latorre} J. Latorre, E. Rico and G. Vidal, Quant. Inf. Comp. \textbf{6}, 48 (2004)
(quant-ph/0304098);
J. Gaite, Proceedings of the conference TH2002, quant-ph/0301120

\bibitem{galindo} A. Galindo and M. A. Mart\'{\i}n-Delgado, Rev. Mod. Phys. {\bf 74}, 347
(2002)

\bibitem{nielsenchuang} E. Schmidt, Math. Ann. {\bf 63}, 433 (1906);
M. Nielsen and I. L. Chuang, {\it Quantum computation and quantum
information} (Cambridge University Press), Cambridge (2000)

\bibitem{bennet} C. H. Bennet, G. Brassard, S. Popescu and B. Schumacher, Phys. Rev. Lett.
{\bf 76}, 722 (1996)

\bibitem{legeza2003} \"O. Legeza and J. S\'olyom, Phys. Rev. B {\bf 68}, 195116 (2003)

\bibitem{callan} C. Callan and F. Wilczek, Phys. Lett. B {\bf 333}, 55 (1994); J. Gaite,
Mod.
Phys. Lett. A {\bf 16}, 1109 (2001)

\bibitem{vidal2002} G. Vidal, J. Latorre, E. Rico and A. Kitaev, Phys. Rev. Lett. {\bf 90},
227902 (2003)
\bibitem{korepin} V. Korepin, Phys. Rev. Lett. {\bf 92}, 96402 (2004)
\bibitem{laflorencie} N. Laflorencie, E. Sorensen, M.-S. Chang and I. Affleck, 
Phys. Rev. Lett. {\bf 96}, 100603 (2006)   

\bibitem{dechiara} G. De Chiara, S. Montenegro, P. Calabrese and R. Fazio, preprint, cond-mat/0512586,
J. Stat. Mech., P03001 (2006)
\bibitem{zhao}  J. Zhao, I. Peschel and X. Wang, Phys. Rev. B 73 024417 (2006)
\bibitem{legezatwosite} \"O. Legeza and J. S\'olyom, preprint, cond-mat/0511081; \"O. Legeza, K. Buchta and J. S\'olyom,
 preprint, cond-mat/0511082

\bibitem{rissler} J. Rissler, R. Noack and S. White, cond-mat/0508524, to appear in Chem. Phys. (2006) 
\bibitem{zanardi} P. Zanardi, Phys. Rev. A {\bf 65}, 42101 (2002); S.-J. Gu, S.-S. Deng, Y.-Q. Lin and H.-Q. Lin, Phys. Rev. Lett.
{\bf 93}, 86402 (2004)
\bibitem{deng} S.-S. Deng, S.-J. Gu and H.-Q. Lin, preprint, cond-mat/0511103

\bibitem{AKLT} I. Affleck, T. Kennedy, E. Lieb and H. Tasaki, Phys. Rev. Lett. {\bf 59}, 799
(1987)

\bibitem{ostlund}  S. \"{O}stlund and S. Rommer, Phys. Rev. Lett. \textbf{75}%
, 3537 (1995);  S. Rommer and S. \"{O}stlund, Phys. Rev. B \textbf{55}, 2164 (1997); M.
Andersson, M. Boman
and S. \"{O}stlund, Phys. Rev. B \textbf{59}, 10493 (1999); H. Takasaki, T.
Hikihara and T. Nishino, J. Phys. Soc. Jpn. \textbf{68}, 1537 (1999)

\bibitem{dukelskyMPS} J. Dukelsky et al, Europhys. Lett. {\bf 43}, 457 (1997)

\bibitem{peschel1}  I. Peschel and M. Kaulke, in Ref. \cite{book}, Chap.
3.1(II)

\bibitem{okunishi1}K. Okunishi, Y. Hieida and Y.Akutsu, Phys. Rev. E \textbf{59}, R6227
(1999)

\bibitem{peschelint}  I. Peschel, M. Kaulke and \"{O}. Legeza, Annalen der
Physik \textbf{8}, 153 (1999),(cond-mat/9810174) %

\bibitem{ors}  \"{O}. Legeza and G. F\'{a}th, Phys. Rev. B \textbf{53},
14349 (1996); M-B Lepetit and G. Pastor, Phys. Rev. B \textbf{58}, 12691
(1998)

\bibitem{chungpeschel} M. C. Chung and I. Peschel, Phys. Rev. B {\bf 64}, 064412 (2001)

\bibitem{boschi} C. Degli Esposti Boschi and F. Ortolani, Eur. Phys. J. B {\bf 41}, 503
(2004)

\bibitem{verstcirac} F. Verstraete and J. I. Cirac, Phys. Rev. B {\bf 73}, 094423 (2006)

\bibitem{chungpeschel2} M. C. Chung and I. Peschel, Phys. Rev. B {\bf 62}, 4191 (2000)

\bibitem{legeza2004} \"{O}. Legeza and J. S\'olyom, Phys. Rev. B \textbf{70}, 205118
(2004)

\bibitem{whitesinglesite} S. White, preprint, Phys. Rev. B {\bf 72}, 180403 (2005)

\bibitem{delgadorg}  M. A. Mart\'{\i}n-Delgado and G. Sierra, Int. J. Mod.
Phys. A, \textbf{11}, 3145 (1996)

\bibitem{sierraqg}  G. Sierra and M. A. Mart\'{\i}n-Delgado, in ''The Exact
Renormalization Group '' by a Krasnitz, R Potting, Y A Kubyshin and P. S. de
Sa (Eds.) World Scientific Pub Co; ISBN: 9810239394, (1999),
(cond-mat/9811170)

\bibitem{peschelcorr} I. Peschel, J. Phys. A: Math. Gen. {\bf 36},
L205 (2003)

\bibitem{sierrairf}  G. Sierra and T. Nishino, Nucl. Phys. B \textbf{495},
505 (1997) (cond-mat/9610221)

\bibitem{wada}  W. Tatsuaki, Phys. Rev. E \textbf{61}, 3199 (2000); T. Wada
and T. Nishino, cond-mat/0103508 (Proceedings of the Conference on
Computational Physics 2000 (CCP2000), Gold Coast, Queensland, Australia, 3-8
December 2000)

\bibitem{ian}  I. P. McCulloch, M. Gulacsi, S. Caprara and A. Juozapavicius,
J. Low Temp. Phys. 117, 323 (1999)

\bibitem{affleck}  E.\ S.\ S\o rensen and I. Affleck, Phys.\ Rev.\ B \textbf{%
51}, 16115 (1995)

\bibitem{simetrias}  I.P. McCulloch and M. Gulacsi, Europhys. Lett. \textbf{57}%
, 852 (2002); I. P. McCulloch, A. R. Bishop and M. Gulacsi, Philos. Mag. B \textbf{81},
1603 (2001)

\bibitem{ramaseshasim}  S. Ramasesha, S. Pati, H. R. Krishnamurthy, Z. Shuai and J. L.
Br\'edas, Phys. Rev. B \textbf{54}, 7598 (1996);
S. Ramasesha, S. Pati, H. R. Krishnamurthy, Z. Shuai and J. L.
Br\'edas, Synth. Metals \textbf{85}, 1019 (1997)

\bibitem{ducroo}  M.S.L. du Croo de Jongh and J.M.J. van Leeuwen, Phys. Rev.
B \textbf{57}, 8494 (1998)


\bibitem{carbon} H-H. Lin et al, preprint, cond-mat/0410654.

\bibitem{karendin}  K. Hallberg, Phys. Rev. B \textbf{52}, 9827 (1995).


\bibitem{white3}  S. R. White and D. Huse, Phys. Rev. B \textbf{48}, 3844
(1993)

\bibitem{sorensen}  E. S. S{\o }rensen and I. Affleck, Phys. Rev. B \textbf{%
49}, 13235 (1994)

\bibitem{sorensen2}  E. S. S{\o }rensen and I. Affleck, Phys. Rev. B \textbf{%
49}, 15771 (1994); E. Polizzi, F. Mila and E. S{\o }rensen, Phys. Rev. B
\textbf{58}, 2407 (1998)

\bibitem{batista}  C. Batista, K. Hallberg and A. Aligia, Phys. Rev. B
\textbf{58}, 9248 (1998) 
\bibitem{batista2}  C. Batista, K. Hallberg and A. Aligia, Phys. Rev. B  
\textbf{60}, 12553 (1999)
\bibitem{batista3} K. Hallberg,  C. Batista and A. Aligia, Physica B,
\textbf{259}, 1017 (1999) 
\bibitem{batista4}E. Jannod, C. Payen, K. Schoumacker, C. Batista,
K. Hallberg and A. Aligia, Phys. Rev B, \textbf{62}, 2998 (2000)

\bibitem{laukamp}  M. Laukamp et al., Phys. Rev. B \textbf{5}, 10755 (1998)

\bibitem{ng}  T-K. Ng, J. Lou and Z. Su, Phys. Rev. B \textbf{61}, 11487
(2000)

\bibitem{affleck3}  J. Lou, S. Qin, T-K. Ng, Z. Su and I. Affleck, Phys.
Rev. B \textbf{62}, 3786 (2000)

\bibitem{marston4}  S-W Tsai and J. B. Marston, Ann. Phys. (Leipzig) 8,
Special Issue, 261 (1999)


\bibitem{sieling}  M. Sieling, U. L\"{o}w, B. Wolf, S. Schmidt, S. Zvyagin
and B.L\"{u}thi, Phys. Rev. B \textbf{61}, 88 (2000)

\bibitem{nishi}  Y. Nishiyama, K. Totsuka, N. Hatano and M. Suzuki, J. Phys.
Soc. Jpn \textbf{64}, 414 (1995)

\bibitem{qinedge}  S. Qin, T. Ng and Z-B Su, Phys. Rev. B \textbf{52}, 12844
(1995)

\bibitem{uli1}  U. Schollw{\"{o}}ck and T. Jolicoeur, Europhys. Lett.
\textbf{30}, 493 (1995)

\bibitem{wangqin}  X. Wang, S. Qin and Lu Yu, Phys. Rev. B. \textbf{60},
14529 (1999)


\bibitem{capone}  M. Capone and S. Caprara, Phys. Rev. B \textbf{64}, 184418
(2001)

\bibitem{yulu2}  J. Lou, X. Dai, S. Qin, Z. Su and L. Yu, Phys. Rev. B
\textbf{60}, 52 (1999)

\bibitem{ercolesi}  E. Ercolessi, G. Morandi, P. Pieri and M. Roncaglia,
Europhys. Lett. \textbf{49}, 434 (2000)

\bibitem{qin}  S. Qin, X. Wang and Lu Yu, Phys. Rev. B \textbf{56}, R14251
(1997)
\bibitem{rizzi} M. Rizzi et al, Phys. Rev. Lett. \textbf{95}, 240404 (2005)
\bibitem{karen12}  K. Hallberg, P. Horsch and G. Mart\'{\i}nez, Phys. Rev. B,
\textbf{52}, R719 (1995)

\bibitem{boos}  H. E. Boos, V. E. Korepin, Y. Nishiyama and M. Shiroishi, J.
Phys. A\textbf{35}, 4443 (2002); V. E. Korepin, S. Lukyanov, Y. Nishiyama
and M. Shiroishi, Phys. Lett. A {\bf 312}, 21 (2003)

\bibitem{shiroishi}  M. Shiroishi, M. Takahashi and Y. Nishiyama, J. Phys.
Soc. Jpn. \textbf{70}, 3535 (2001)

\bibitem{hiki}  T. Hikihara and A. Furusaki, Phys. Rev. B \textbf{58}, R583
(1998)

\bibitem{luttinger}  G. Fath, Phys. Rev. B {\bf 68}, 134445 (2003)

\bibitem{lou2}  J. Lou, S. Qin, C. Chen, Z. Su and L. Yu, Phys. Rev. B
\textbf{65}, 064420 (2002)

\bibitem{lou3} J. Lou, S. Qin and C. Chen,  Phys. Rev. Lett. {\bf 91}, 087204
(2003)

\bibitem{caprara}  F. Capraro and C. Gros, Eur. Phys. J. B {\bf 29}, 35 (2002)

\bibitem{hieida}  Y. Hieida, K. Okunishi and Y. Akutsu, Phys. Rev. B \textbf{%
64}, 224422 (2001)

\bibitem{byrnes}  T. Byrnes, R. Bursill, H-P. Eckle, C. Hamer and A.
Sandvik, Phys. Rev. B{\bf 66}, 195313 (2002)

\bibitem{marston2}  S-W. Tsai and J. B. Marston, Phys. Rev. B \textbf{62},
5546 (2000)

\bibitem{pastor}  M-B. Lepetit, M. Cousy and G. Pastor, Eur. Phys. J. B
\textbf{13}, 421 (2000); H. Otsuka, Phys. Rev. B \textbf{53}, 14004 (1996)

\bibitem{zhuo} W. Zhuo, X. Wang and Y. Wang, preprint, cond-mat/0501693


\bibitem{karen32}  K. Hallberg, X. Wang, P. Horsch and A. Moreo, Phys. Rev.
Lett. \textbf{76}, 4955 (1996)

\bibitem{adriana}  A. Moreo, Phys. Rev. B \textbf{35}, 8562 (1987); T. Ziman
and H. Schulz, Phys. Rev. Lett \textbf{59}, 140 (1987)

\bibitem{yamashita}  Y. Yamashita, N. Shibata and K. Ueda, Phys. Rev. B
\textbf{58}, 9114 (1998);  Y. Yamashita, N. Shibata and K. Ueda, J. Phys. Soc. Jpn.
{\bf 69}, 242 (2000);  Y. Yamashita, N. Shibata and K. Ueda, Phys. Rev. B \textbf{61},
4012 (2000)

\bibitem{onu}  A. Onufriev and B. Marston, Phys. Rev. B \textbf{59}, 12573
(1999)

\bibitem{affleck2}  C. Itoi, S. Qin and I. Affleck, Phys. Rev. B \textbf{61}%
, 6747 (2000)

\bibitem{haas2}  W. Yu and S. Haas, Phys. Rev. B \textbf{63}, 024423 (2001)
%
%

\bibitem{bursill}  R.\ J.\ Bursill, T.\ Xiang and G.\ A.\ Gehring, J.\
Phys.\ A \textbf{28} 2109 (1994)

\bibitem{dim1}  R.\ J.\ Bursill, G.\ A.\ Gehring, D.\ J.\ J.\ Farnell, J.\
B.\ Parkinson, T. Xiang and C. Zeng, J.\ Phys.\ C \textbf{7} 8605 (1995)

\bibitem{dim2}  U.\ Schollw\"{o}ck, Th.\ Jolicoeur and T.\ Garel, Phys.\
Rev.\ B \textbf{53}, 3304 (1996)

\bibitem{dim3}  R.\ Chitra, S.\ Pati, H.\ R.\ Krishnamurthy, D.\ Sen and S.\
Ramasesha, Phys.\ Rev.\ B \textbf{52}, 6581 (1995); S.\ Pati, R.\ Chitra,
D.\ Sen, H.\ R.\ Krishnamurthy and S.\ Ramasesha, Europhys.\ Lett.\ \textbf{%
33}, 707 (1996); J. Malek, S. Drechsler, G. Paasch and K. Hallberg, Phys.
Rev. B \textbf{56}, R8467 (1997); E. S{\o }rensen et al in Ref.\cite{book},
Chap.1.2 (Part II) and references therein; D. Augier, E. S{\o }rensen, J.
Riera and D. Poilblanc, Phys. Rev. B \textbf{60}, 1075 (1999)

\bibitem{dim4}  Y.\ Kato and A.\ Tanaka, J.\ Phys.\ Soc.\ Jpn.\ \textbf{63},
1277 (1994)

\bibitem{dim5}  S.\ R.\ White and I.\ Affleck, Phys. Rev. B \textbf{54},
9862 (1996)

\bibitem{dim6}  G. Bouzerar, A. Kampf and G. Japaridze, Phys. Rev. B \textbf{%
58}, 3117 (1998); G. Bouzerar, A. Kampf and F. Sch\"{o}nfeld,
cond-mat/9701176, unpublished; M-B. Lepetit and G. Pastor, Phys. Rev. B
\textbf{56}, 4447 (1997)

\bibitem{kaburagi}  M. Kaburagi, H. Kawamura and T. Hikihara, J. Phys. Soc.
Jpn. \textbf{68}. 3185 (1999)

\bibitem{hikihara2}  T. Hikihara, J. Phys. Soc. Jpn. \textbf{71}, 319
(2002); T. Hikihara, M. Kaburagi, H. Kawamura and T. Tonegawa, Phys. Rev. B
\textbf{63}, 174430 (2001)

\bibitem{patialt}  S. Pati, S. Ramasesha and D. Sen, Phys. Rev. B \textbf{55}%
, 8894 (1996); J. Phys. Cond. Matt. \textbf{9}, 8707 (1997); T. Tonegawa,
T. Hikihara, T. Nishino, M. Kaburagi, S. Miyashita, H.J. Mikeska, J. Mag.
\textbf{177-181}, 647 (1998)

\bibitem{noackchain}  M.\ Azzouz, L.\ Chen and S.\ Moukouri, Phys.\ Rev.\ B
\textbf{50} 6223 (1994); S.\ R.\ White, R.\ M.\ Noack and D.\ J.\ Scalapino,
Phys.\ Rev.\ Lett.\ \textbf{73} 886 (1994); K.\ Hida, J.\ Phys.\ Soc.\ Jpn.\
\textbf{64} 4896 (1995); T.\ Narushima, T.\ Nakamura and S.\ Takada, J.\
Phys.\ Soc.\ Jpn.\ \textbf{64} 4322 (1995); U.\ Schollw\"{o}ck and D.\ Ko,
Phys.\ Rev.\ B \textbf{53} 240 (1996); G. Sierra, M. A. Mart\'{\i}n-Delgado,
S. White and J. Dukelsky, Phys. Rev. B \textbf{59}, 7973 (1999); S. White,
Phys. Rev. B \textbf{53}, 52 (1996)

\bibitem{kawaguchi}  A. Kawaguchi, A. Koga, K. Okunishi and N. Kawakami,
Phys. Rev. B {\bf 65}, 214405 (2002)

\bibitem{trumper}  A. Trumper and C. Gazza, Phys. Rev. B \textbf{64}, 134408
(2001)

\bibitem{roger}  M. Roger, Phys. Rev. B \textbf{63}, 144433 (2001)

\bibitem{hikihara3}  T. Hikihara and A. Furusaki, Phys. Rev. B \textbf{63},
134438 (2001)
\bibitem{capriotti} L. Capriotti, D. Scalapino and S. White, Phys. Rev. Lett. \textbf{93},
177004 (2004)
\bibitem{rodriguezlaguna} M. A. Mart\'{\i}n-Delgado, J. Rodriguez-Laguna and G. Sierra, 
Phys. Rev. B 72 104435 (2005)

\bibitem{schmid}  A. Laeuchli, G. Schmid and M. Troyer, Phys. Rev. B {\bf 67}, 100409
(2003)

\bibitem{hikihara}  T. Hikihara, T. Momoi and X. Hu, Phys. Rev. Lett. {\bf 90}, 087204
(2003)

\bibitem{nunner}  T. Nunner, P. Brune, T. Kopp, M. Windt and M. Gr\"{u}ninger, Phys. Rev.
B {\bf 66}, 180404 (2002)

\bibitem{honda}  Y. Honda and T. Horiguchi, preprint, cond-mat/0106426

\bibitem{patising}  S. K. Pati and R. R. P. Singh, Phys. Rev. B \textbf{60},
7695 (1999); S. R. White and R. Singh, Phys. Rev. Lett. \textbf{85}, 3330
(2000)

\bibitem{maeshima}  N. Maeshima, M. Hagiwara, Y. Narumi, K. Kindo, T. C.
Kobayashi and K. Okunishi,Journal of Physics: Cond. Matt. {\bf 15}, 3607 (2003);
N. Maeshima and K. Okunishi, Phys. Rev. B \textbf{62}, 934-939 (2000)

\bibitem{itoi}  C. Itoi and S. Qin, Phys. Rev. B \textbf{63}, 224423 (2001)

\bibitem{yulu}  J. Lou, J. Dai, S. Qin, Z. Su and L. Yu, Phys. Rev. B
\textbf{62}, 8600 (2000)

\bibitem{triang}  C. Raghu, I. Rudra, S. Ramasesha and D. Sen, Phys. Rev. B
\textbf{62}, 9484 (2000)

\bibitem{schotte}  A. Honecker, M. Kaulke and K.D. Schotte, Eur. Phys. J. B
\textbf{15}, 423 (2000)

\bibitem{gendiar3}  A. Gendiar and A. Surda, Phys. Rev. B \textbf{62}, 3960 (2000)
\bibitem{otsuka} H. Otsuka, Y. Okabe an K. Okunishi, Phys. Rev. E 73 035105(R) (2006)
\bibitem{fouet} J.-B. Fouet et al, Phys. Rev. B {\bf 73}, 014409 (2006)


\bibitem{tandon}  K. Tandon, S. Lal, S. K. Pati, S. Ramasesha and D. Sen, Phys.
Rev. B \textbf{59}, 396 (1999); R. Citro, E. Orignac, N. Andrei, C. Itoi and S. Qin, J.
Phys.: Cond. Mat. \textbf{12}, 3041 (2000)

\bibitem{lou}  J. Lou, C. Chen and S. Qin, Phys. Rev. B \textbf{64}, 144403
(2001)

\bibitem{wanglu}  X. Wang and L. Yu, Phys. Rev. Lett. \textbf{84}, 5399
(2000)

\bibitem{silva} J. Silva-Valencia and E. Miranda, Phys. Rev. B {\bf 65}, 024443
(2002); J. Silva-Valencia, J. C. Xavier and E. Miranda, Phys. Rev. B 71, 024405 (2005)
\bibitem{gu} B. Gu, G. Su and S. Gao, J. Phys: Cond. Matt.{\bf 17}, 6081 (2005)  

\bibitem{patisen}  S. K. Pati, S. Ramasesha and D. Sen, in \textit{%
Magnetism: Molecules to Materials IV}, eds. J.S. Miller and M. Drillon
(Wiley-VCH Weinheim, 2002), Chapter 4, (cond-mat/0106621)

\bibitem{noackhubb}  L.\ Chen and S.\ Moukouri, Phys.\ Rev.\ B \textbf{53}
1866 (1996); S.\ J.\ Qin, S.\ D.\ Liang, Z.\ B.\ Su and L.\ Yu, Phys.\ Rev.\
B \textbf{52} R5475 (1995); R. Noack in Ref.\cite{book}, Chap.1.3 (Part II);
S. Daul and R. Noack, Phys. Rev. B \textbf{61}, 12361 (2000); M. Vojta, R.
Hetzel and R. Noack, Phys. Rev. B \textbf{60}, R8417 (1999)

\bibitem{penc}  K. Penc, K. Hallberg, F. Mila and H. Shiba, Phys. Rev. Lett.
\textbf{77}, 1390 (1996); Phys. Rev. B, \textbf{55}, 15475 (1997)

\bibitem{bulut}  N. Bulut, Advances in Physics \textbf{51}, 1587 (2002)

\bibitem{malvezzi}  A. Malvezzi, T. Paiva and R. dos Santos, Phys. Rev. B {\bf 66}, 064430
(2002)

\bibitem{eric} E. Jeckelmann, Phys. Rev. Lett. {\bf 89}, 236401 (2002); G. P. Zhang, Phys.
Rev. B {\bf 68}, 153101 (2003) and erratum Phys. Rev. B {\bf 69}, 199902(E) (2004); see also
comment on this paper by E. Jeckelmann, Phys. Rev. B {\bf 71}, 197101 (2005) and the reply
by G. P. Zhang, Phys. Rev. B {\bf 71}, 197102 (2005). 


\bibitem{zhang2}  Y. Z. Zhang, C. Wu and H. Q.Lin, Phys. Rev. B \textbf{65},
115101 (2002); Y. Z. Zhang, C. Q. Wu and H. Q. Lin, Phys. Rev. B {\bf 72}, 125126 (2005);
 Y. Z. Zhang, Phys. Rev. Lett. {\bf 92}, 246404 (2004)

\bibitem{arno}  A. P. Kampf, M. Sekania, G. I. Japaridze, Ph. Brune, J. Phys.: Condens.
Matter {\bf 15} 5895 (2003)

\bibitem{baeriswyl}  C. Aebischer, D. Baeriswyl and R. M. Noack, Phys. Rev.
Lett. \textbf{86}, 468 (2001)

\bibitem{aoki}  R. Arita and H. Aoki, Phys. Rev. B \textbf{61}, 12261
(2000); R. Arita, Y. Shimoi, K. Kuroki and H. Aoki, Phys. Rev. B \textbf{57}%
, 10609 (1998)

\bibitem{daul2}  S. Daul, Eur. J. Phys. B {\bf 14}, 649 (2000)

\bibitem{qin2}  S. Qin, J. Lou, T. Xiang, G-S. Tian and Z. Su, Phys. Rev. B {\bf 68},
045110 (2003)

\bibitem{armando}  A. Aligia, K. Hallberg, C. Batista and G. Ortiz, Phys.
Rev. B \textbf{61}, 7883 (2000)
\bibitem{weisse} A. Weisse, R. Bursill, C. Hamer and Z. Weihong, preprint, cond-mat/0511528

\bibitem{kurt}  V. Meden, W. Metzner, U. Schollw\"ock, O. Schneider, T.
Stauber and K. Schoenhammer, , Europhys. J B \textbf{16}, 631 (2000)

\bibitem{falicov}  P. Farkasovsky, Phys. Rev. B \textbf{65}, 081102 (2002)

\bibitem{schuster}  C. Schuster, R. Roemer and M. Schreiber, Phys. Rev. B
\textbf{65}, 115114 (2002)

\bibitem{fibonacci}  K. Hida, Phys. Rev. Lett. \textbf{86}, 1331 (2001)
\bibitem{ejima} S. Ejima, F. Gebhard and S. Nishimoto, Europhys. Lett. \textbf{70}, 492 
(2005) 
\bibitem{phi4}  W. Lay and J. Rudnick, Phys. Rev. Lett. \textbf{88}, 057203
(2002); Y. Nishiyama, J. Phys. A\textbf{34}, 11215 (2001); S. G. Chung, Phys.
Rev. E \textbf{62}, 3262 (2000)

\bibitem{srinivasan}  B. Srinivasan and M. B. Lepetit, Phys. Rev. B \textbf{%
66}, 024421 (2002); H. Sakamoto, T. Momoi and K. Kubo, Phys. Rev. B \textbf{%
65}, 224403 (2002)
\bibitem{xavier2} J. Xavier, H. Onishi, T. Hotta and E. Dagotto, 
Phys. Rev. B \textbf{73}, 014405 (2006) 
\bibitem{hozoi} L. Hozoi and S. Nishimoto, preprint, cond-mat/0512219

\bibitem{hubbardladder}  K. Hamacher, C. Gros and W. Wenzel, Phys. Rev.
Lett. \textbf{88}, 217203 (2002)

\bibitem{vojta}  M. Vojta, A. Huebsch and R. M. Noack, Phys. Rev. B \textbf{%
63}, 045105 (2001); Z. Weihong, J. Oitmaa, C.J. Hamer and R.J. Bursill, J.
Phys. C \textbf{13}, 433 (2001)

\bibitem{liang2}  Youngho Park, S. Liang and T. K. Lee, Phys. Rev. B \textbf{%
59}, 2587 (1999)
\bibitem{ramasesha3} Y. Yan, S. Mazumdar and S. Ramasesha, preprint, cond-mat/0601481   
\bibitem{edegger2} B. Edegger, H. Evertz and R. Noack, preprint, cond-mat/0510325
\bibitem{scalapino2}  S. Rommer, S. R. White and D. J. Scalapino, Phys. Rev.
B \textbf{61}, 13424 (2000)

\bibitem{hager} G. Hager et al, preprint, Phys. Rev. B {\bf 71}, 075108 (2005)

\bibitem{fjaerestad} J. O. Fjaerestad, J. B. Marston and U. Schollw\"ock , Ann. Phys. (NY) 321 894 
(2006), cond-mat/0412709
\bibitem{edegger} B. Edegger, H. G. Evertz and R. M. Noack, Phys. Rev. B \textbf{72}, 085131 (2005) 
\bibitem{roux} G. Roux et al., Phys. Rev. B {\bf 72}, 014523 (2005)

\bibitem{fermion}  R. M. Noack, S. R. White and D. J. Scalapino, Phys. Rev.
Lett. \textbf{73}, 882 (1994); S. R. White, R. M. Noack and D. J. Scalapino,
J. Low Temp. Phys. \textbf{99}, 593 (1995); R. M. Noack, S. R. White and D.
J. Scalapino, Europhys. Lett. \textbf{30}, 163 (1995); C. A. Hayward, D.
Poilblanc, R. M. Noack, D. J. Scalapino and W. Hanke, Phys. Rev. Lett.
\textbf{75}, 926 (1995); S. White and D. Scalapino, Phys. Rev. Lett. \textbf{%
81}, 3227 (1998); E. Jeckelmann, D. Scalapino and S. White, Phys. Rev. B
\textbf{58}, 9492 (1998)

\bibitem{nishimoto}  S. Nishimoto, E. Jeckelmann and D. Scalapino, Phys. Rev. B {\bf 66},
245109 (2002)

\bibitem{marston}  J. B. Marston, J. O. Fjaerestad and A. Sudbo, Phys. Rev.
Lett. \textbf{89}, 056404 (2002)

\bibitem{time}  U. Schollw\"ock, S. Chakravarty, J. O. Fjaerestad, J. B.
Marston and M. Troyer, Phys. Rev. Lett. {\bf 90}, 186401 (2003)

\bibitem{bourbonnais}  L. G. Caron and C. Bourbonnais, Phys. Rev. B \textbf{%
66}, 045101 (2002)

\bibitem{ziosi}  G. Fano, F. Ortolani, A. Parola and L. Ziosi, Phys. Rev. B
\textbf{60}, 15654 (1999)

\bibitem{landau}  N. Shibata and D. Yoshioka, Phys. Rev. Lett. \textbf{86},
5755 (2001)
\bibitem{landau1} N. Shibata and D. Yoshioka, Physica E \textbf{12}, 43 (2002)
\bibitem{landau2} N. Shibata and D. Yoshioka, J. Phys. Soc. Jpn. 
\textbf{75}, 043712 (2006) 

\bibitem{rings}  V. Meden and U. Schollw\"{o}ck, Phys. Rev. B {\bf 67}, 035106 (2003);
F. Carvalho Dias, I. Pimentel and M. Henkel, Phys. Rev. B
\textbf{73}, 075109 (2006) 

\bibitem{teo}  T.\ A.\ Costi, P.\ Schmitteckert, J.\ Kroha and P.\ W\"{o}%
lfle, Phys.\ Rev.\ Lett.\ \textbf{73}, 1275 (1994); S.\ Eggert and I.\
Affleck, Phys.\ Rev.\ Lett.\ \textbf{75}, 934 (1995); E.\ S.\ S\o rensen and
I. Affleck, Phys.\ Rev.\ B \textbf{51}. 16115 (1995); X.\ Q.\ Wang and S.\
Mallwitz, Phys.\ Rev.\ B \textbf{53}, R492 (1996); W.\ Wang, S.\ J.\ Qin,
Z.\ Y.\ Lu, L. Yu and Z.\ B.\ Su, Phys.\ Rev.\ B \textbf{53}, 40 (1996); C.\
C.\ Yu and M.\ Guerrero, Phys. Rev. B \textbf{54}, 15917 (1996); A. Furusaki
and T. Hikihara, Phys. Rev. B \textbf{58}, 5529 (1998)

\bibitem{egger}  K. Hallberg and R. Egger, Phys. Rev. B \textbf{55}, 8646
(1997)

\bibitem{impchains}  C. Schuster and U. Eckern, Ann. Phys. (Leipzig) 11, 901-915 (2002);
W. Zhang, J. Igarashi and P.
Fulde, Phys. Rev. B \textbf{56}, 654 (1997)

\bibitem{meden}  V. Meden, W. Metzner, U. Schollw\"ock and K. Schoenhammer,
Phys. Rev. B \textbf{65}, 045318 (2002)
\bibitem{andergassen} S. Andergassen et al, Phys. Rev. B \textbf{73}, 045125
(2006)

\bibitem{kondoins}  C.\ C.\ Yu and S.\ R.\ White, Phys.\ Rev.\ Lett.\
\textbf{71}, 3866 (1993); C.\ C.\ Yu and S.\ R.\ White, Physica B \textbf{199%
}, 454 (1994)

\bibitem{kondolatt}  S.\ Moukouri and L.\ G.\ Caron, Phys.\ Rev.\ B \textbf{%
52}, 15723 (1995); N.\ Shibata, T.\ Nishino, K.\ Ueda and C.\ Ishii, Phys.\
Rev.\ B \textbf{53}, R8828 (1996); M.\ Guerrero and R.\ M.\ Noack, Phys.\
Rev.\ B \textbf{53}, 3707 (1996)

\bibitem{kondoneck}  H.\ Otsuka and T.\ Nishino, Phys.\ Rev.\ B \textbf{52},
15066 (1995); S.\ Moukouri, L.\ G.\ Caron, C.\ Bourbonnais and L.\ Hubert,
Phys.\ Rev.\ B \textbf{51}, 15920 (1995)

\bibitem{kondoxavier}  J. C. Xavier, E. Novais and E. Miranda, Phys. Rev. B
\textbf{65}, 214406 (2002)

\bibitem{kondonecklace}  T. Yamamoto, R. Manago and Y. Mori, J. Phys. Soc. Jpn. {\bf 72},
3204
(2003)

\bibitem{kondojuoza}  I. P. McCulloch, A. Juozapavicius, A. Rosengren and M.
Gulacsi, Phys. Rev. B \textbf{65}, 052410 (2002)

\bibitem{kondoguerrero}  M. Guerrero and R.M. Noack, Phys. Rev. B \textbf{63}%
, 144423 (2001)

\bibitem{watanabe}  S. Watanabe, J. Phys. Soc. Jpn. \textbf{69}, 2947 (2000)

\bibitem{xavier}  J. Xavier, R. Pereira, E. Miranda and I. Affleck,
Phys. Rev. Lett. {\bf 90}, 247204 (2003)
\bibitem{masa} S. Watanabe, M. Imada and K. Miyake, J. Phys. Soc. Jpn. 75 043710 (2006)

\bibitem{wata}  S. Watanabe, Y. Kuramoto, T. Nishino and N. Shibata, J.
Phys. Soc. Jpn \textbf{68}, 159 (1999)

\bibitem{kondomoreno}  J. Moreno, S. Qin, P. Coleman and L. Yu, Phys. Rev. B
\textbf{64}, 085116 (2001)
\bibitem{schauerte} T. Schauerte et al, Phys. Rev. Lett. {\bf 94}, 147201 (2005)

\bibitem{chen}  S.\ Moukouri, L.\ Chen and L.\ G.\ Caron, Phys.\ Rev.\ B
\textbf{53}, R488 (1996)

\bibitem{riera}  J. Riera, K. Hallberg and E. Dagotto, Phys. Rev. Lett.
\textbf{79}, 713 (1997); E. Dagotto et al., Phys. Rev. B \textbf{58}, 6414
(1998)

\bibitem{kondodaniel}  D. Garc\'{\i}a, K. Hallberg, C. Batista, M. Avignon
and B. Alascio, Phys. Rev Lett., \textbf{85}, 3720 (2000); D. J. Garc\'{\i}a,
K. Hallberg, C. D. Batista, S. Capponi, D. Poilblanc, M. Avignon and B.
Alascio, Phys. Rev. B. \textbf{65}, 134444 (2002)

\bibitem{imada}  B. Ammon and M. Imada, J. Phys. Soc. Jpn. \textbf{70}, 547
(2001) %


\bibitem{neuber} D. Neuber et al, Phys. Rev. B {\bf 73}, 014401 (2006)


\bibitem{daniel}  D. Garc\'{\i}a, K. Hallberg, B. Alascio and M. Avignon, Physical Rev. Lett., 
\textbf{93}, 17, 177204 (2004)

\bibitem{mcculloch}I.P. McCulloch, A. Juozapavicius, A. Rosengren and M. Gulacsi,
Phys. Rev. B, \textbf{65}, 52410 (2002)


\bibitem{hida}  K. Hida, J. Phys. Soc. Jpn. \textbf{65}, 895 (1996) and 3412
(1996) (erratum); J. Phys. Soc. Jpn. \textbf{66}, 330 (1997); J. Phys. Soc.
Jpn. \textbf{66}, 3237 (1997); Prog. Theor. Phys. Suppl. \textbf{145}, 320
(2002); K. Hida, preprint, cond-mat/0602016
\bibitem{lajko} P. Lajk\'o, E. Carlon, H. Rieger and F. Igl\'oi, Phys. Rev. B \textbf{72}, 094205
(2005)
\bibitem{hida1}  K. Hida, J. Phys. Soc. Jpn. \textbf{68}, 3177 (1999)

\bibitem{hida2}  K. Hida, Phys. Rev. Lett. \textbf{83}, 3297 (1999)

\bibitem{griffiths}  F. Igloi, R. Juhasz and P. Lajko, Phys. Rev. Lett.
\textbf{86}, 1343 (2001)

\bibitem{spinladders}  R. M\'{e}lin, Y-C. Lin, P. Lajk\'{o}, H. Rieger and
F. Igl\'{o}i, Phys. Rev. B \textbf{65}, 104415 (2002)

\bibitem{schmitt}  P. Schmitteckert, T. Schulze, C. Schuster, P. Schwab and
U. Eckern, Phys. Rev. Lett. \textbf{80}, 560 (1998); P. Schmitteckert and U.
Eckern, Phys. Rev. B \textbf{53}, 15397 (1996)

\bibitem{schmitt2}  D. Weinmann, P. Schmitteckert, R. Jalabert and J.
Pichard, Eur. Phys. J. B \textbf{19}, 139-156 (2001)

\bibitem{jala}  P. Schmitteckert, R. Jalabert, D. Weinmann and J. L.
Pichard, Phys. Rev. Lett. \textbf{81}, 2308 (1998); E. Gambetti, Phys. Rev. B {\bf 72}, 165338 (2005)

\bibitem{carlonigloi}  E. Carlon, P. Lajko and F. Igl\'{o}i, Phys. Rev.
Lett. \textbf{87}, 277201 (2001) %

\bibitem{maruyama} I. Maruyama, N. Shibata and K. Ueda, J. Phys. Soc. Jpn. {\bf 73}, 3239 (2004);
{\it ibid} J. Phys. Soc. Jpn. {\bf 73}, 434 (2004)

\bibitem{sade} M. Sade et al., Phys. Rev. B {\bf 71}, 153301 (2005)

\bibitem{berkovits} R. Berkovits, F. von Oppen and Y. Gefen, Phys. Rev. Lett. {\bf 
94}, 076802 (2005)

\bibitem{ye} F. Ye etal, Phys. Rev. B {\bf 72}, 233409 (2005)
\bibitem{chernyshev} A. L. Chernyshev, A. H. Castro Neto and S. White, Phys. Rev. 
Lett. {\bf 94}, 036407 (2005)

\bibitem{paredes}, B. Paredes et al, Nature {\bf  429}, 277 (2004); T. Kinoshita, T. Wenger
and D. S. Weiss, Science {\bf 305}, 1125 (2004)
\bibitem{schmidt} B. Schmidt, L. Plimak and M. Fleischhauer, Phys. Rev. A {\bf 71}, 041601 (2005)
\bibitem{deng2} X.-L Deng, D. Porras and J. Cirac, Phys. Rev. B {\bf 72}, 075351 (2005)

\bibitem{oka1} T. Oka and N. Nagaosa, Phys. Rev. Lett. \textbf{95}, 266409 (2005) 


\bibitem{moukouri2}  L. Caron and S. Moukouri, Phys. Rev. Lett. \textbf{76},
4050 (1996);  L. Caron and S. Moukouri, Phys. Rev. B \textbf{56}, R8471 (1997)

\bibitem{jeck}  E. Jeckelmann and S. White, Phys. Rev. B \textbf{57}, 6376
(1998)

\bibitem{noackberlin}  R. Noack, S. White and D. Scalapino in \textit{%
Computer Simulations in Condensed Matter Physics VII}, edited by D. Landau,
K.-K. Mon and H.-B. Sch\"{u}ttler (Springer Verlag, Heidelberg and Berlin,
1994)

\bibitem{jeckbook}  E. Jeckelmann, C. Zhang and S. White in Ref. \cite{book}%
, Chap. 5.1 (II)
\bibitem{jeckreview} E. Jeckelmann and H. Fehske, preprint, Proc. of the "Enrico Fermi" Int. School of 
Physics, Varenna, Italy, June 2005  cond-mat/0510637

\bibitem{tezuka} M. Tezuka, R. Arita and H. Aoki, Phys. Rev. Lett. \textbf{95}, 226401 (2005);
ibid. Physica B \textbf{359-361}, 708 (2005) 

\bibitem{jeck2}  C. Zhang, E. Jeckelmann and S. White, Phys. Rev. Lett.
\textbf{80}, 2661 (1998); E. Jeckelmann, C. Zhang and S. R. White, Physical
Review B \textbf{60}, 7950 (1999)
\bibitem{matsuedaholstein} H. Matsueda, T. Tohyama and S. Maekawa, preprint, cond-mat/0511068
  
\bibitem{nishiyama3}  Y. Nishiyama,  Phys. Rev. B {\bf 64}, 064510 (2001)

\bibitem{nishiyama5} Y. Nishiyama, Eur. Phys. J. B \textbf{12}, 547 (1999)

\bibitem{barfordpeierls} W. Barford and R. J. Bursill, Phys. Rev. Lett. {\bf 95}, 137207 (2005)

\bibitem{bursillphonon}  R. Bursill, Y. McKenzie and C. Hammer, Phys. Rev.
Lett. \textbf{80}, 5607 (1998);  R. Bursill, Y. McKenzie and C. Hammer, Phys.
Rev. Lett. \textbf{83}, 408 (1999); R. Bursill, Phys. Rev. B \textbf{60}, 1643 (1999) %

\bibitem{krish}  R. Pai, R. Pandit, H. Krishnamurthy and S. Ramasesha, Phys.
Rev. Lett. \textbf{76}, 2937 (1996) (see also the comment by N. V. Prokof'ev
and B. V. Svistunov, Phys. Rev. Lett. \textbf{80}, 4355 (1998)); S. Rapsch,
U. Schollw\"{o}ck and W. Zwerger, Europhys. Lett. \textbf{46}, 559 (1999)

\bibitem{monien}  T. K\"{u}hner and H. Monien, Phys. Rev. B \textbf{58},
R14741 (1998)

\bibitem{chung}  I. Peschel and M-C. Chung, J. Phys. A: Math. Gen. \textbf{32%
}, 8419 (1999)

\bibitem{maurel1}  P. Maurel and M-B. Lepetit, Phys. Rev. B \textbf{62},
10744 (2000); P. Maurel, M-B. Lepetit and D. Poilblanc, Eur. Phys. J.
\textbf{B 21}, 481 (2001)

\bibitem{xiang2}  T. Xiang, Phys. Rev. B \textbf{53}, R10445 (1996)

\bibitem{nishimoto3}  S. Nishimoto, E. Jeckelmann, F. Gebhard and R. Noack,
Phys. Rev. B \textbf{65}, 165114 (2002)

\bibitem{porras} D. Porras, F. Verstraete and J. I. Cirac, Phys. Rev. B {\bf 73}, 014410 (2006)



\bibitem{duksierra}  J. Dukelsky and G. Sierra, Phys. Rev. Lett. \textbf{83}%
, 172 (1999); ibid. Phys. Rev. B {\bf 61}, 12302 (2000); J. Dukelsky and S. Pittel,
Phys. Rev. C {\bf 63}, R061303 (2001)

\bibitem{dukelskyreview}  J. Dukelsky and S. Pittel, Rep. Prog. Phys.
\textbf{67}, 513 (2004); S. Pittel et al., Rev. Mex. Fis. {\bf 49S4}, 82 (2003)

\bibitem{gobert} D. Gobert, M. Schechter, U. Schollw\"ock and J. Von Delft, Phys. Rev.
Lett. {\bf 93}, 186402 (2004);  D. Gobert, U. Schollw\"ock and J. Von
Delft, Eur. Phys. J. B  {\bf 38}, 501 (2004)

\bibitem{dukelsky}  J. Dukelsky and G. Dussel, Phys. Rev. C \textbf{59},
R3005 (1999);  J. Dukelsky and S. Pittel, Phys. Rev.  C {\bf 63}
R061303 (2001); J. Dukelsky, S. Pittel, S. S. Dimitrova and M. V. Stoitsov, Phys.
Rev. C {\bf 65}, 054319 (2002)

\bibitem{papenbrock} T. Papenbrock and D. J. Dean, J. Phys. G {\bf 31}, S1377 (2005)
 
\bibitem{rotureau} J. Rotureau et al., preprint, cond-mat/0603021

\bibitem{delgadoqcd}  M. A. Mart\'{\i}n-Delgado and G. Sierra, Phys. Rev.
Lett. \textbf{83}, 1514 (1999)

\bibitem{wilson2}  S. Glazek and K. Wilson, Phys. Rev. D \textbf{48}, 5863
(1993); \textbf{49}, 4214 (1994)

\bibitem{dlcq}  T. Eller, H.-C. Pauli and S. Brodsky, Phys. Rev. D \textbf{35%
}, 1493 (1987)


\bibitem{ppp}  G. Fano, F. Ortolani and L. Ziosi, J. Chem. Phys. \textbf{108}%
, 9246 (1998),(cond-mat/9803071); R. Bursill and W. Barford, Phys. Rev.
Lett. \textbf{82}, 1514 (1999); ibid. preprint, cond-mat/0512649; E. Moore, W. Barford and R. Bursill, 
Phys. Rev. B {\bf 71}, 115107 (2005); 
A. Race, W. Barford and R. J. Bursill, Phys. Rev. B {\bf 67}, 245202
(2003)

\bibitem{pppreview} For a review see S. Ramasesha, S.K. Pati, Z. Shuai and J. L. Br'edas, 
Adv. Quantum Chem. 38, 121 (2000)

\bibitem{keplerate}  M. Exler and J. Schnack, Phys. Rev. B {\bf 67}, 094440 (2003)

\bibitem{bruce}  B. Normand, X. Wang, X. Zotos and D. Loss, Phys. Rev. B
\textbf{63}, 184409 (2001)

\bibitem{timm} C. Timm and U. Schollw\"ock, Phys. Rev. B {\bf 71}, 224414 (2005)

\bibitem{polyacenes}  C. Raghu, Y. Anusooya Pati and S. Ramasesha,
Journal-ref: J. Phys. A \textbf{34}, 11215 (2001)

\bibitem{barford}  W. Barford and R. Bursill, Chem. Phys. Lett. \textbf{268}%
, 535(1997); W. Barford, R. Bursill and M. Lavrentiev, J. Phys: Cond. Matt,
\textbf{10}, 6429 (1998); W. Barford in Ref.\cite{book}, Chap.2.3 (Part II)
and references therein; M. Lavrentiev, W. Barford, S. Martin, H. Daly, R.
Bursill, Physical Review B \textbf{59}, 9987 (1999); R Bursill and W. Barford, 
Phys. Rev. B {\bf 66}, 205112 (2002)

\bibitem{reptons}  E. Carlon, A. Drzewinski and J. van Leeuwen, J. Chem.
Phys. \textbf{117}, 2425 (2002); Phys. Rev. E \textbf{64}, 010801(R) (2001);
M. Paessens and G. Sch\"{u}tz, Phys. Rev. E \textbf{66}, 021806 (2002);

\bibitem{dendrimers}  M. A. Mart\'{\i}n-Delgado, J. Rodriguez-Laguna and G.
Sierra, Phys. Rev. B \textbf{65}, 155116 (2002)

\bibitem{whitemol}  S. White and R. Martin, J. Chem. Phys. \textbf{110},
4127 (1999)

\bibitem{whiteorth}  S. White in Ref.~\cite{book}, Chap. 2.1.

\bibitem{legezaqc}  \"{O}. Legeza, J. R{\"{o}}der and B.A. Hess, Phys. Rev. B {\bf 67},
125114 (2003); ibid., Mol. Phys. {\bf 101}, 2019 (2003)

\bibitem{daulwhite}  S. Daul, I. Ciofini, C. Daul and S. R. White,
Int. J. Quantum Chem. {\bf 79}, 331 (2000)

\bibitem{bau}  C. Bauschlicher and P. Taylor, J. Chem. Phys. \textbf{85},
2779 (1986)

\bibitem{mitrushenko} A. O. Mitrushenkov et al, J. Chem Phys. {\bf 115}, 6815 (2001)

\bibitem{chan} G. K-L. Chan and M. Head-Gordon, J. Chem Phys. {\bf 116}, 4462 (2002)


\bibitem{liangpang}  S. Liang and H. Pang, Europhys. Lett. \textbf{32}, 173
(1995); ibid. Phys. Rev. B {\bf 49}, 9214 (1994)

\bibitem{w1}  S. White and D. Scalapino, Phys. Rev. B \textbf{55}, 14701
(1997)

\bibitem{w2}  S. White and D. Scalapino, Phys. Rev. B \textbf{55}, 6504
(1997)

\bibitem{w3}  S. White and D. Scalapino, Phys. Rev. B \textbf{57}, 3031
(1998)

\bibitem{arrigoni}  E. Arrigoni, A. P. Harju, W. Hanke, B. Brendel and S.A.
Kivelson, Phys. Rev. B \textbf{65}, 134503 (2002); J. Bonca, J. E.
Gubernatis, M. Guerrero, E. Jeckelmann and S. R. White, Phys. Rev. B \textbf{%
61}, 3251 (2000); A. L. Chernyshev, S. White and A. H. Castro Neto, Phys.
Rev. B, \textbf{65}, 214527 (2002); S. White and D. Scalapino, Phys. Rev. B
\textbf{61}, 6320 (2000); {\it ibid} Phys. Rev. B {\bf 70}, 220506 (2004)

\bibitem{w4}  S. White and D. Scalapino, Phys. Rev. Lett. \textbf{80}, 1272
(1998); {\it ibid.} Phys. Rev. B {\bf 61}, 6320 (2000)

\bibitem{bishop}  I. P. McCulloch, A. R. Bishop and M. Gulacsi, Phil. Mag. B
\textbf{81}, 1603 (2001)

\bibitem{white5}  S. R. White and I. Affleck, Phys. Rev. B \textbf{64},
024411 (2001)
\bibitem{weng} M. Q. Weng, et al, preprint, cond-mat/0508186

\bibitem{su}  T. Xiang, J. Lou and Z. Su, Phys. Rev. B \textbf{64}, 104414
(2001)

\bibitem{nishino5}  N. Maeshima, Y. Hieida, Y. Akutsu, T. Nishino and K.
Okunishi, Phys. Rev. E \textbf{64} (2001) 016705

\bibitem{sierra3d}  M.A. Martin-Delgado, J. Rodriguez-Laguna and G. Sierra,
Nucl. Phys. B \textbf{601}, 569 (2001)

\bibitem{henelius}  P. Henelius, Phys. Rev. B \textbf{60}, 9561 (1999)
\bibitem{farnell} D. J. J. Farnell, Phys. Rev. B {\bf 68}, 134419 (2003)

\bibitem{moukouri5}  S. Moukouri and L. G. Caron, Phys. Rev. B {\bf 67}, 092405 (2003); J. V.
Alvarez and S. Moukouri, Int. J. Mod. Phys. C {\bf 16}, 843 (2005); S. Moukouri, J. Stat. Mech. P02002 
(2006); S. Moukouri, Phys. Rev. B {\bf 70}, 014403 (2004)

\bibitem{shibatareview}  N. Shibata, J. Phys. A, \textbf{36} 381 (2003)

\bibitem{verstuli} F. Verstraete et al, preprint, cond-mat/0504305

\bibitem{ulirevt} U. Schollw\"ock, J. Phys. Soc. Jpn. {\bf 74} (Suppl), 246 (2005); 
see also S. Manmana, A. Muramatsu and R. Noack, AIP Conf. Proc. {\bf 789}, 269 (2005)
(cond-mat/0502396)

\bibitem{cazalilla} M. A. Cazalilla and J. B. Marston, Phys. Rev. Lett. {\bf 88}, 256403
(2002)

\bibitem{crank} W. H. Press et al, Numerical Recipes in C++, Cambridge Univ. Press, 1993,
second edition.

\bibitem{daley} A. Daley, C. Kollath, U. Scholl\"ock and G. Vidal, J. Stat. Mech.: Theor.
Exp. P04005 (2004)


\bibitem{luo} H. G. Luo, T. Xiang and X. Q. Wang, Phys. Rev. Lett. {\bf 91}, 049701
(2003)(comment to \cite{cazalilla}); M. A. Cazalilla and J. B. Marston, Phys. Rev. Lett.
{\bf 91}, 049702 (2003) (response to previous comment)

\bibitem{schmitteckert} P. Schmitteckert, Phys. Rev. B {\bf 70}, 121302 (2004)
\bibitem{schmitteckert2} G. Schneider and P. Schmitteckert, preprint, cond-mat/0601389.

\bibitem{feiguin2} A. Feiguin and S. White, Phys. Rev. B {\bf 72}, 020404 (2005) 

\bibitem{feiguin} S. White and A. Feiguin, Phys. Rev. Lett. {\bf 93}, 076401 (2004)

\bibitem{vidal} G. Vidal, Phys. Rev. Lett. {\bf 91}, 147902 (2003); G. Vidal, Phys. Rev.
Lett. {\bf 93}, 040502 (2004)

\bibitem{gobert2} D. Gobert et al, Phys. Rev. E 71 036102 (2005)

\bibitem{kollath} C. Kollath et al, Phys. Rev. A {\bf 71}, 053606 (2005)

\bibitem{clark} S. R. Clark and D. Jaksch, Physical Review A,{\bf 70}, 043612 (2004)
  
\bibitem{ulisep} C. Kollath, U. Schollw\"ock and W. Zwerger, Phys. Rev. Lett. {\bf 95}, 176401 (2005)

\bibitem{jagla} E. A. Jagla, K. Hallberg and C. A. Balseiro, Phys. Rev. B {\bf 47}, 5849
(1993)

\bibitem{oka} T. Oka and H. Aoki, Phys. Rev. Lett. {\bf 95}, 137601 (2005)

\bibitem{alhassanieh} K.A. Al-Hassanieh et al, preprint cond-mat/0601411 
\bibitem{micheli} A. Micheli, a. j. Daley, D. Jaksch and P. Zoller, Phys. Rev. Lett. {\bf 93}, 
140408 (2004)

\bibitem{kuehner}  T. K\"{u}hner and S. White, Phys. Rev. B \textbf{60},
335(1999)

\bibitem{ramasesha}  Y. Anusooya, S. Pati and S. Ramasesha, J. Chem. Phys.
\textbf{106}, 1 (1997); S. Ramasesha, K. Tandon, Y. Anusooya and S. Pati,
Proc. of SPIE, \textbf{3145}, 282 (1997); S. Ramasesha, Z. Shuai and J. Br%
\'{e}das, Chem. Phys. Lett. \textbf{245}, 224 (1995)


\bibitem{jeckelmann}  E. Jeckelmann, Phys. Rev. B \textbf{66}, 045114 (2002)

\bibitem{carlos}  E. R. Gagliano and C. A. Balseiro, Phys. Rev. Lett.
\textbf{59}, 2999 (1987).

\bibitem{proj}  G. Grosso and G. Partori Parravicini, in \textit{Memory
Function Approaches to Stochastic Problems in Condensed Matter}, Adv. in
Chemical Physics, \textbf{62}, 133 (Wiley, N. Y., 1985)

\bibitem{linden}  P. Horsch and W. von der Linden, Z. Phys. B \textbf{72}
181 (1981)

\bibitem{nishiyama}  Y. Nishiyama, Eur. J. Phys. B \textbf{12}, 547 (1999)

\bibitem{zhangjeck}  C. L. Zhang, E. Jeckelmann and S. White, Phys. Rev. B
\textbf{60}, 14092 (1999)

\bibitem{yuhaas}  W. Q. Yu and S. Haas, Phys. Rev. B, \textbf{63}, 024423
(2000)

\bibitem{okunishi}  K. Okunishi, Y. Akutsu, N. Akutsu and T. Yamamoto, Phys. Rev. B
\textbf{64}, 104432 (2001)

\bibitem{kancharla}  S. S. Kancharla and C. J. Bolech, Phys. Rev. B \textbf{%
64}, 085119 (2001)

\bibitem{matsueda} H. Matsueda et al, Phys. Rev. B {\bf 72}, 075136 (2005)

\bibitem{monien2} T. Hand, J. Kroha and H. Monien, preprint, cond-mat/0602352

\bibitem{essler}  F. Essler, F. Gebhard and E. Jeckelmann, Phys. Rev. B
\textbf{64}, 125119 (2001); E. Jeckelmann, F. Gebhard and F. Essler, Phys.
Rev. Lett. \textbf{85}, 3910 (2000)

\bibitem{jeck03} E. Jeckelmann, Phys. Rev. B {\bf 67}, 075106 (2003)

\bibitem{kim}  Y.-J Kim et al, 
Phys. Rev. Lett. \textbf{92}, 137402 (2004); J. Rissler, F. Gebhard and E. Jeckelmann, J.
Phys.: Cond. Matt. {\bf 17}, 4093 (2005)

\bibitem{benthien} H. Benthien, F. Gebhard and E. Jeckelmann, Phys. Rev. Lett. {\bf 92},
256401 (2004)

\bibitem{nishimotojeck} S. Nishimoto and E. Jeckelmann, J. Phys.: Cond. Matt. {\bf 16}, 613
(2004)
\bibitem{nishimoto2imp} S. Nishimoto, T. Pruschke and R. Noack, J. Phys.: Cond. Matt 18 981 (2006)

\bibitem{pati}  S. K. Pati and R. Singh, Phys. Rev. B \textbf{60}, 7695 (1999)

\bibitem{kuehner2}  T. D. K\"{u}hner, S. White and H. Monien, Phys. Rev. B
\textbf{61}, 12474 (2000)

\bibitem{raas}  C. Raas, G. Uhrig and F. B. Anders, Phys. Rev. B \textbf{69}%
, 041102(R) (2004); C. Raas and G. Uhrig, Eur. J. Phys. B{\bf 45}(3), 293 (2005)

\bibitem{wolfle} D. Bohr, P. Schmitteckert and P. W\"olfle, Europhys. Lett. \textbf{73},
246 (2006)

\bibitem{hofstetter}  W. Hofstetter, Phys. Rev. Lett. \textbf{85}, 1508
(2000)


\bibitem{nishino}  T. Nishino, J. Phys. Soc. Jpn. \textbf{64}, 3598 (1995);
see also T. Nishino in Ref. \cite{book}, Chap. 5(I); T. Nishino and K.
Okunishi in \textit{Strongly Correlated Magnetic and Superconducting Systems}%
, Ed. G. Sierra and M. A. Mart\'{\i}n-Delgado (Springer, Berlin, 1997)

\bibitem{clasico}  H. Trotter, Proc. Am. Math. Soc. \textbf{10}, 545 (1959);
M. Suzuki, Prog. Theor. Phys. \textbf{56}, 1454 (1976); R. Feynman and A.
Hibbs \textit{Quantum Mechanics and Path Integrals} (McGraw-Hill, 1965)

\bibitem{bursill96}  R. Bursill, T. Xiang and G. Gehring, J. Phys. Cond.
Mat. \textbf{8}, L583 (1996); X. Q. Wang and T. Xiang, Phys. Rev. B \textbf{%
56}, 5061 (1997); N. Shibata, J. Phys. Soc. Jpn \textbf{66} 2221 (1997)

\bibitem{gendiar2}  A. Gendiar and A. Surda, Phys. Rev. B \textbf{63},
014401 (2001)

\bibitem{shibata2}  T. Nishino and N. Shibata, J. Phys. Soc. Jpn. \textbf{68}%
, 3501 (1999); T. Enss and U. Schollw\"ock, J. Phys. A \textbf{34}, 7769
(2001)

\bibitem{stochastic}  A. Kemper, A. Schadschneider and J. Zittartz, J. Phys.
A: Math. Gen. \textbf{34}, L279 (2001); T. Enss and U. Schollw\"{o}ck, J
Phys A: Math. Gen. \textbf{34}, 7769 (2001)

\bibitem{baxter}  R. Baxter, J. Math. Phys. \textbf{9}, 650 (1968); J. Stat
Phys. \textbf{19}, 461 (1978)

\bibitem{okunishi3}  T. Nishino and K. Okunishi, J. Phys. Soc. Jpn. \textbf{65%
},891 (1996); T. Nishino and K. Okunishi, J. Phys. Soc. Jpn. \textbf{66}, 3040 (1997); T.
Nishino, K.  Okunishi and M. Kikuchi, Phys. Lett. A \textbf{213}, 69 (1996)

\bibitem{takasaki}  H. Takasaki, T. Nishino and Y. Hieida, J. Phys. Soc.
Jpn. \textbf{70}, 1429 (2001)

\bibitem{nishino3}  T. Nishino, Y. Hieida, K. Okunishi, N. Maeshima, Y.
Akutsu and A. Gendiar, Prog. Theor. Phys. \textbf{105}, 409 (2001)

\bibitem{ritter}  C. Ritter and G. von Gehlen, in ''Quantization, Gauge
Theory and Strings'', ed. A.Semikhatov et al., Vol. I, p.563-578, Scientific
World Pub. Co. (2001) (cond-mat/0009255)

\bibitem{ueda} K. Ueda et al, J. Phys. Soc. Jpn. Suppl. 74 111 (2005);
ibid. J. Phys. Soc. Jpn. 74 1871 (2005);
K. Ueda et al, J. Phys. Soc. Jpn. 75 014003 (2006)

\bibitem{ueda2}  N. Shibata and K. Ueda, J. Phys. Soc. Jpn. Vol.70, 3690
(2001)

\bibitem{okunishi2}  T. Nishino and K. Okunishi, J. Phys. Soc. Jpn. \textbf{%
67}, 3066 (1998); T. Nishino, K. Okunishi, Y. Hieida, N. Maeshima and Y.
Akutsu, Nucl. Phys. B \textbf{575}, 504 (2000)
\bibitem{okunishi4} K. Okunishi, J. Phys. Soc. Jpn. \textbf{74}, 3186 (2005)

\bibitem{drz}  E. Carlon and A. Drzewi\'{n}ski, Phys. Rev. Lett. \textbf{79}, 1591 (1997);
E. Carlon and A. Drzewi\'{n}ski, Phys. Rev. E \textbf{57}, 2626 (1998); E. Carlon, A.
Drzewi\'{n}ski and J. Rogiers, Phys. Rev. B \textbf{58}, 5070 (1998);
A. Drzewi\'{n}ski, A. Ciach and A. Maciolek, Eur. Phys. J. B \textbf{5}, 825 (1998);
A. Maciolek, A. Ciach and A. Drzewi\'{n}ski, Phys. Rev. E
\textbf{60}, 2887 (1999)

\bibitem{drz2}  A. Drzewi\'{n}ski, A. Maciolek and R. Evans, Phys. Rev. Lett.
\textbf{85}, 3079 (2000); A. Drzewi\'{n}ski, Phys. Rev. E \textbf{62}, 4378
(2000)

\bibitem{kaulke}  M.-C. Chung, M. Kaulke, I. Peschel, M. Pleimling and W.
Selke, Eur. Phys. J. B \textbf{18}, 655 (2000)

\bibitem{carlon}  E. Carlon and F. Igl\'{o}i, Phys. Rev. B \textbf{57}, 7877
(1998); F. Igl\'{o}i and E. Carlon, Phys. Rev. B \textbf{59}, 3783 (1999); E. Carlon, C.
Chatelain and B. Berche, Phys. Rev. B \textbf{60}, 12974 (1999)

\bibitem{kondev}  J. Kondev and J. Marston, Nucl. Phys. B \textbf{497}, 639
(1997); T. Senthil, B. Marston and M. Fisher, Phys. Rev. B \textbf{60}, 4245
(1999); J. Marston and S. Tsai, Phys. Rev. Lett. \textbf{82}, 4906 (1999);
S. Tsai and J. Marston, Ann. Phys. (Leipzig) \textbf{8}, Special Issue,
261(1999)

\bibitem{peschel}  M. Kaulke and I. Peschel, Eur. Phys. J. B \textbf{5}, 727
(1998)

\bibitem{hieida2}  Y. Hieida, J. Phys. Soc. Jpn. \textbf{67}, 369 (1998)

\bibitem{carlon1}  E. Carlon, M. Henkel and U. Schollw\"{o}ck, Eur. Phys. J.
B \textbf{12}, 99(1999); E. Carlon, M. Henkel and U. Schollw\"ock, Phys. Rev.
E \textbf{63}, 036101 (2001)

\bibitem{jef}  J. Hooyberghs, E. Carlon and C. Vanderzande, Phys. Rev. E
\textbf{62}, 036124 (2001)

\bibitem{chung3}  S.G. Chung, Phys. Rev. B \textbf{60}, 11761(1999) %
\bibitem{nishiyama4} Y. Nishiyama, Phys. Rev. E,\textbf{72}, 036104 (2005); ibid. 
Phys. Rev. E 66, 061907 (2002)
and Phys. Rev. E 68, 31901 (2003)


\bibitem{moukouri}  S. Moukouri and L. Caron, Phys. Rev. Lett. \textbf{77},
4640 (1996).

\bibitem{moukouri3}  S. Moukouri and L. Caron, see Ref.\cite{book}, Chap.
4.5(II)

\bibitem{moukouri4}  S. Moukouri, preprint, cond-mat/0011169

\bibitem{zhang}  W. Zhang, J. Igarashi and P. Fulde, J. Phys. Soc. Jpn.
\textbf{66}, 1912 (1997) 


\bibitem{bursill2}  R. Bursill, T. Xiang and G. Gehring, J. Phys. C \textbf{8%
}, L583 (1996)

\bibitem{trotter}  H. Trotter, Proc. Am. Math. Soc. \textbf{10}, 545 (1959);
M. Suzuki, Prog. Theor. Phys. \textbf{56}, 1454 (1976)

\bibitem{xiangwang}  X. Wang and T. Xiang, Phys. Rev. B \textbf{56}, 5061
(1997)

\bibitem{shibata}  N. Shibata, J. Phys. Soc. Jpn. \textbf{66}, 2221 (1997)

\bibitem{xiangt}  T. Xiang, Phys. Rev. B \textbf{58}, 9142 (1998)

\bibitem{ammon1}  N. Shibata, B. Ammon, T. Troyer, M. Sigrist and K. Ueda,
J. Phys. Soc. Jpn. \textbf{67}, 1086 (1998)

\bibitem{maisinger}  K. Maisinger, U. Schollw\"{o}ck, S. Brehmer, H-J. Mikeska
and S. Yamamoto, Phys. Rev. B \textbf{58}, R5908 (1998)

\bibitem{rommer}  S. Rommer and S. Eggert, Phys. Rev. B \textbf{59}, 6301
(1999)

\bibitem{maruyama1}  I. Maruyama, N. Shibata and K. Ueda, Phys. Rev. B
\textbf{65}, 174421 (2002)

\bibitem{maisinger2}  K. Maisinger and U. Schollw\"{o}ck, Phys. Rev. Lett.
\textbf{81}, 445 (1999)

\bibitem{ammon}  B. Ammon, M. Troyer, T. Rice and N. Shibata, Phys. Rev.
Lett. \textbf{82}, 3855 (1999); N. Shibata and H. Tsunetsugu, J. Phys. Soc.
Jpn. \textbf{68}, 3138 (1999)

\bibitem{naefwang}  F. Naef and X. Wang, Phys. Rev. Lett. \textbf{84}, 1320
(2000)

\bibitem{klumper}  A. Kl\"{u}mper, R. Raupach and F. Sch\"{o}nfeld, Phys.
Rev. B \textbf{59}, 3612 (1999)

\bibitem{sirker} J. Sirker and A. Kl\"umper, Phys. Rev. B {\bf 66}, 245102 (2002)

\bibitem{ohta} Y. Ohta, T. Nakaegawa and S. Ejima, Phys. Rev. B {\bf 73}, 045101  
(2006)

\bibitem{verstripoll} F. Verstraete, J. J. Garc\'{\i}a Ripoll and J. I. Cirac, Phys. Rev.
Lett. {\bf 93}, 207204 (2004)

\bibitem{zwolak} M. Zwolak and G. Vidal,  Phys. Rev. Lett. \textbf{93}, 207205 (2004)
\bibitem{feiguintemp} A. Feiguin and S. White, Phys. Rev. B \textbf{72}, 220401 (2005)

\bibitem{wangbook} X. Wang, K. Hallberg and F. Naef in Ref.\cite{book},
Chap.7(I)

\bibitem{mutou} T. Mutou, N. Shibata and K. Ueda, Phys. Rev. Lett {\bf
81}, 4939 (1998) (erratum: Phys. Rev. Lett. {\bf 82}, 3727 (1999))

\bibitem{naef} F. Naef, X. Wang, X. Zotos and W. von der Linden,
Phys. Rev. B {\bf 60}, 359 (1999)

\bibitem{sirker2} J. Sirker and A. Kl\"umper, Phys. Rev. B {\bf
71}, 241101 (2005)  



\bibitem{pt}  G. Kotliar and D. Vollhardt, \textsl{Physics Today} \textbf{57}, 53 (2004)

\bibitem{hf}  J.E. Hirsch and R.M. Fye, \textsl{Phys. Rev. Lett.} \textbf{56}%
, 2521 (1986)

\bibitem{mr}  M.J. Rozenberg, \textsl{Phys. Rev. B} \textbf{55}, R4855 (1997)


\bibitem{bulla}  R. Bulla, \textsl{Phys. Rev. Lett.} \textbf{83}, 136 (1999)

\bibitem{vollhardt}  R. Bulla, T. Costi and D. Vollhardt, \textsl{Phys. Rev.
B} \textbf{64}, 045103 (2001)

\bibitem{garciadmft}  D. Garc\'{\i}a, K. Hallberg and M. Rozenberg, Phys. Rev. Lett. {\bf
93}, 246403 (2004)

\bibitem{review}  A. Georges, G. Kotliar, W. Krauth and M. J. Rozenberg
\textsl{Rev. Mod. Phys.} \textbf{68}, 13 (1996)

\bibitem{qimiao}  Q. Si, M. J. Rozenberg, G. Kotliar, and A. E. Ruckenstein
\textsl{Phys. Rev. Lett.} \textbf{72}, 2761 (1994)

\bibitem{mpl}  M. J. Rozenberg, G. Moeller and G. Kotliar, \textsl{Mod.
Phys. Lett. B} \textbf{8}, 535 (1994)

\bibitem{gebhard}  See F. Gebhard, E. Jeckelmann, S. Mahlert, S. Nishimoto and R. Noack,
 \textsl{Eur. Phys. J B}
\textbf{36}, 491 (2003), for a previous attempt using a particular algorithm
of the DMRG restricted to the small U metallic regime.

\bibitem{jeckelmandmft}  S. Nishimoto, F. Gebhard and E. Jeckelmann, J.
Phys.: Cond. Matt. \textbf{16}, 7063, (2004)

\bibitem{gk}  A. Georges and G. Kotliar, \textsl{Phys. Rev. B} \textbf{45},
6479 (1992)

\bibitem{zrg}  X. Y. Zhang, M. J. Rozenberg and G. Kotliar, \textsl{Phys.
Rev. Lett.} \textbf{70}, 1666 (1993), A. Georges and W. Krauth Phys. Rev. B 48, 7167-7182
(1993).

\bibitem{garciamiranda} D. Garcia,  E. Miranda, K. Hallberg and M. Rozenberg, preprint, cond-mat/0608248 

\bibitem{alps} F. Alet et al, preprint cond-mat/0410407. http://alps.comp-phys.org/



\end{thebibliography}
\end{document}